\documentclass[]{mosi}

\usepackage{helvet}

\usepackage{amsmath} 
\usepackage{natbib}
\usepackage{graphicx}
\usepackage{subcaption} 

\usepackage[toc,page,header]{appendix}
\usepackage[utf8]{inputenc} 
\usepackage[T1]{fontenc}    
\usepackage{hyperref}       
\usepackage{url}            
\usepackage{booktabs}       

\usepackage{amsfonts}       
\usepackage{nicefrac}       
\usepackage{microtype}      
\usepackage{wrapfig}
\usepackage{amssymb}        

\usepackage{titletoc}
\usepackage{minitoc}

\usepackage{array}
\usepackage{etoolbox}

\definecolor{lightblue}{RGB}{200, 230, 255}  
\definecolor{headerblue}{RGB}{150, 200, 255} 

\usepackage{pgfplots}
\usepackage{xcolor}
\usepackage{float}
\usepackage{comment}
\usepackage{multirow}
\usepackage{makecell}
\usepackage{siunitx}
\usepackage{tikz}
\usepackage{pgf-pie}
\usepackage[export]{adjustbox}

\usepackage{ragged2e}
\usepackage{tabularx}
\usepackage{caption}
\usepackage{enumitem}
\usepackage{pifont}
\usepackage[hang,flushmargin]{footmisc}

\usepackage{tcolorbox}
\tcbuselibrary{breakable}
\tcbuselibrary{skins}

\usepackage{listings}

\usepackage{threeparttable}
\usepackage{setspace}

\usepackage[scheme=plain]{ctex}

\usepackage[table]{xcolor}

\newcommand{\up}{$\uparrow$}
\newcommand{\down}{$\downarrow$}
\newcommand{\bnum}[1]{\textbf{#1}}

\definecolor{oursgray}{gray}{0.95}

\definecolor{MossCyan}{HTML}{82D9FF} 
\definecolor{MossBlue}{HTML}{82B1FF}

\definecolor{tickG}{HTML}{00C853}
\definecolor{crossR}{HTML}{FF1744}

\newcommand{\cmark}{\textcolor{tickG}{\bfseries\ding{52}}}
\newcommand{\xmark}{\textcolor{crossR}{\bfseries\ding{56}}}
\newcommand{\faHome}{\raisebox{-0.2ex}{\includegraphics[height=2.0ex]{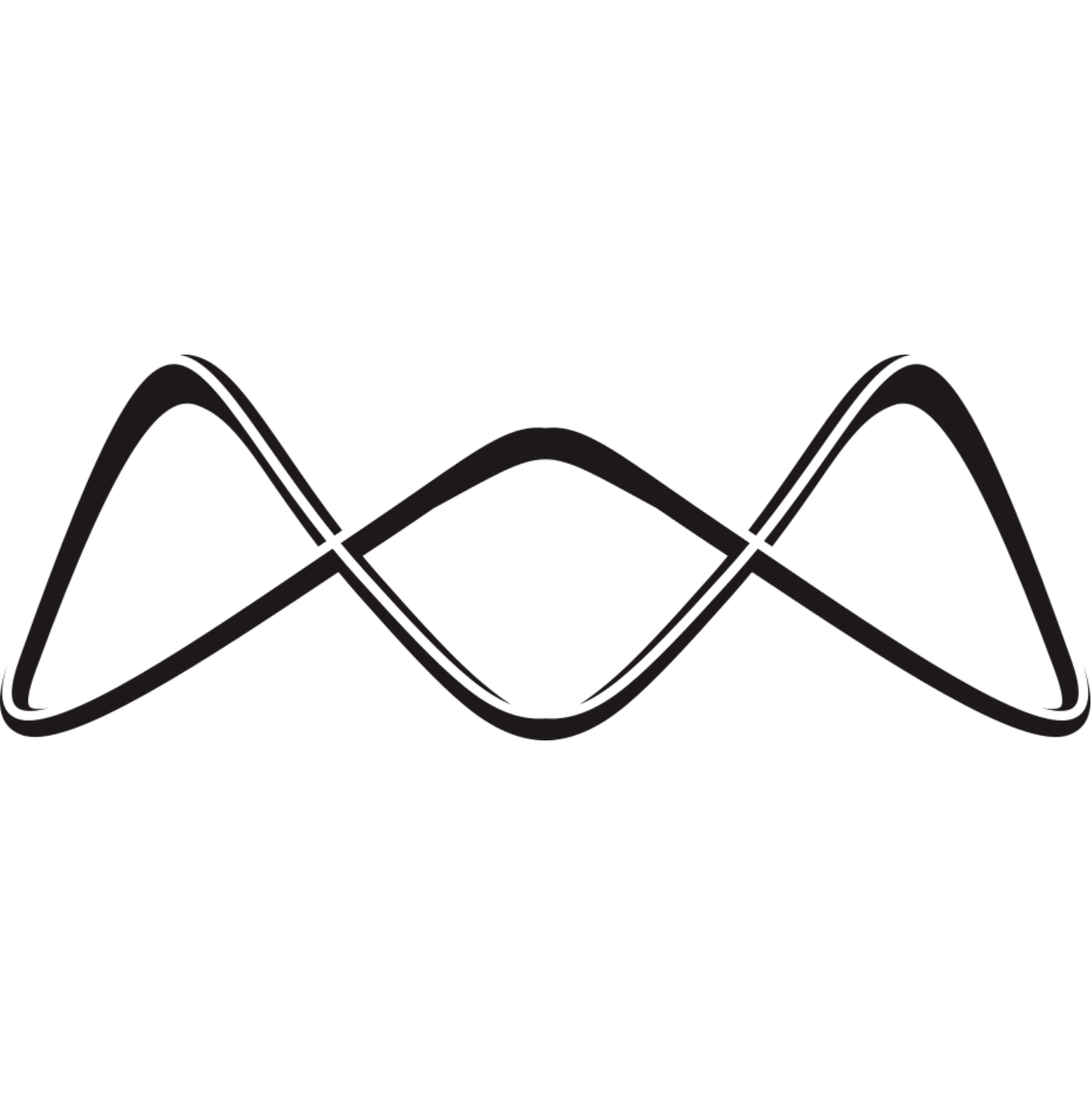}}}
\newcommand{\faPlayCircle}{\raisebox{-0.2ex}{\includegraphics[height=2.0ex]{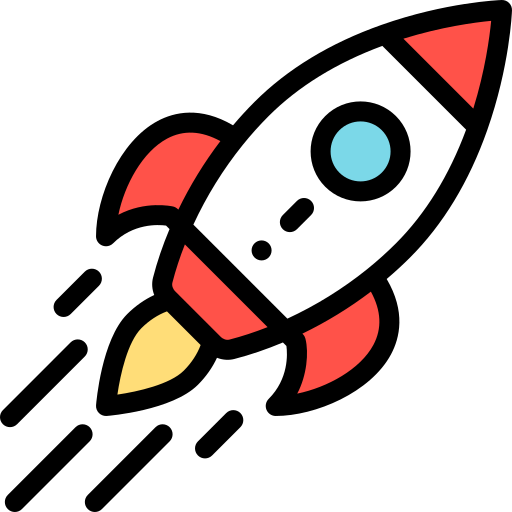}}}
\newcommand{\faCogs}{\raisebox{-0.2ex}{\includegraphics[height=2.0ex]{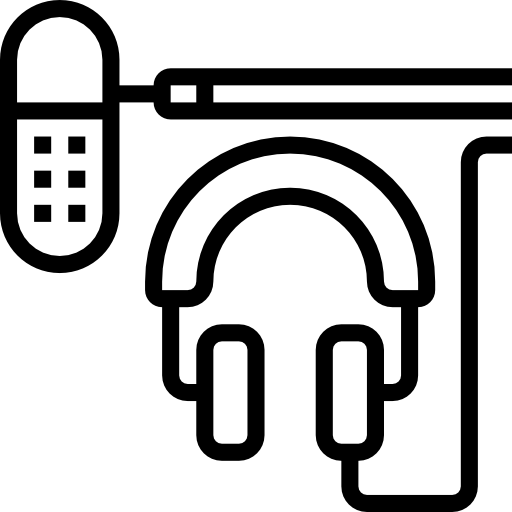}}}
\newcommand{\hflogo}{\raisebox{-0.2ex}{\includegraphics[height=2.0ex]{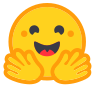}}}
\newcommand{\faGithub}{\raisebox{-0.2ex}{\includegraphics[height=2.0ex]{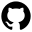}}}

\newtcolorbox{promptbox}[2][]{
    colback=white,
    coltext=black,
    arc=3mm,
    boxrule=0.5pt,
    colframe=black!60!white,
    title={#2},
    colbacktitle=black,
    coltitle=white,
    fonttitle=\bfseries,
    top=8pt,
    bottom=8pt,
    left=10pt,
    right=10pt,
    breakable,
    before upper={%
        \linespread{1}\selectfont
        \setlength{\parskip}{1ex plus 0.2ex minus 0.2ex}%
        \setlength{\parindent}{0pt}%
    },
    #1
}
\title{MOSS-TTS Technical Report}

\author{SII-OpenMOSS Team\textsuperscript{*}}

\abstract{
This technical report presents \textbf{MOSS-TTS}, a speech generation foundation model built on a scalable recipe: discrete audio tokens + autoregressive modeling + large-scale pretraining. Built on \textbf{MOSS-Audio-Tokenizer}, a causal Transformer tokenizer that compresses 24\,kHz audio to 12.5\,fps with variable-bitrate RVQ and unified semantic--acoustic representations, we release two complementary generators: \textbf{MOSS-TTS}, which emphasizes structural simplicity, scalability, and long-context/control-oriented deployment, and \textbf{MOSS-TTS-Local-Transformer}, which introduces a frame-local autoregressive module for higher modeling efficiency, stronger speaker preservation, and a shorter time to first audio. Across multilingual and open-domain settings, MOSS-TTS supports zero-shot voice cloning, token-level duration control, phoneme-/pinyin-level pronunciation control, smooth code-switching, and stable long-form generation. This report summarizes the design, training recipe, and empirical characteristics of the released models.
}

\checkdata[\raisebox{-0.1ex}{\faHome}\hspace{0.4em}Homepage]{\url{https://mosi.cn/models/moss-tts}}
\checkdata[\raisebox{-0.1ex}{\faPlayCircle}\hspace{0.4em}Online Demo]{\url{https://huggingface.co/spaces/OpenMOSS-Team/MOSS-TTS}}
\checkdata[\raisebox{-0.1ex}{\faCogs}\hspace{0.4em}AI Studio]{\url{https://studio.mosi.cn/voice-synthesis}}
\checkdata[\raisebox{-0.1ex}{\hflogo}\hspace{0.4em}Hugging Face]{\url{https://huggingface.co/collections/OpenMOSS-Team/moss-tts}}
\checkdata[\raisebox{-0.1ex}{\faGithub}\hspace{0.4em}GitHub]{\url{https://github.com/OpenMOSS/MOSS-TTS}}

\begin{document}
\maketitle
\begingroup
\renewcommand{\thefootnote}{\fnsymbol{footnote}}
\setcounter{footnote}{1}
\footnotetext{Full contributors can be found in the Contributors section.}
\endgroup



\section{Introduction}

Text-to-speech (TTS) has evolved from task-specific pipelines into a broader paradigm of speech generation that is expected to behave like a foundation model: it should generalize across speakers, languages, speaking styles, and acoustic conditions; support controllable and low-latency synthesis; and remain stable over long-form content \citep{hu2026qwen3,du2025cosyvoice,anastassiou2024seed,xie2025fireredtts}. Recent progress increasingly resembles the scaling trajectory of large language models, where model capacity and data scale unlock emergent capabilities beyond narrow benchmarks \citep{kaplan2020scaling,henighan2020scaling}.

At the same time, scaling speech generation is not a simple matter of ``bigger models.'' Modern approaches must reconcile competing requirements in representation learning and pretraining: (i) the discrete token representation must be compact enough for efficient sequence modeling yet expressive enough to preserve both semantic content and fine-grained acoustics; (ii) the generative model must remain stable over long sequences while staying compatible with streaming constraints; and (iii) the training signal must scale across diverse, noisy, real-world data without relying on brittle cascaded supervision. Much of the recent literature addresses these tensions by introducing multiple intermediate targets, external semantic teachers, refinement stages, or post-hoc alignment.
Such designs can be effective, but they often complicate scaling because each additional module introduces a new supervision contract, new failure modes, and new latency budgets \citep{zhang2023speechtokenizer,defossez2024moshi,du2024cosyvoice,wang2024maskgct,du2025cosyvoice}. 

This report argues for a return to the core of speech generation: learn a high-quality audio tokenizer, train an autoregressive (AR) model over its tokens, and pretrain at scale. Concretely, we pursue the recipe
discrete tokens + AR modeling + large-scale pretraining,
and show that it provides a clean and scalable path to strong quality and controllability in practice. The key intuition is that a sufficiently capable tokenizer turns speech generation into a token prediction problem with a single, universal modeling objective---much like language modeling---thereby making it easier to scale data, compute, and downstream capabilities without continuously expanding the model stack.

MOSS-TTS combines three core components.
\textbf{(1) A high-quality audio tokenizer.}
We build on \textbf{MOSS-Audio-Tokenizer} \citep{gong2026moss}, a causal Transformer-based discrete tokenizer designed for large-scale AR modeling.
It supports variable-bitrate residual vector quantization (RVQ), compressing 24\,kHz audio to 12.5\,fps and enabling streaming-friendly, frame-level encoding and decoding, while preserving high-fidelity reconstruction and semantically informative tokens.
Unlike approaches that depend on external pretrained audio encoders or multi-stage distillation \citep{hsu2021hubert,radford2023robust,zhang2023speechtokenizer,defossez2024moshi,ye2025codec}, MOSS-Audio-Tokenizer is trained end-to-end to jointly optimize acoustic reconstruction and semantic alignment, aiming to maximize scalability and minimize inherited bottlenecks.

\textbf{(2) Large-scale, high-quality pretraining data.}
We build a large-scale, high-quality data pipeline that converts raw open-domain recordings into trainable single-speaker assets with cross-consistency gating (speaker consistency, language consistency, and transcript validity). The resulting corpus spans millions of hours in total, with the majority consisting of carefully filtered multilingual TTS-style supervision and targeted supplements for voice cloning and controllability. This data-centric foundation is essential for robustness across domains (podcasts, audiobooks, broadcast \& news, film \& drama, commentary, and online content) and for multilingual and code-switching behavior.

\textbf{(3) Refined discrete-token modeling for speech generation.}
On top of the tokenizer, we study and deploy discrete AR modeling strategies that remain efficient and stable for long-form synthesis. To serve both research reproducibility and practical deployment constraints, we explore two architectures with explicit tradeoffs. The Delay-Pattern model (\textbf{MOSS-TTS}) uses a single Transformer backbone with multiple prediction heads and an RVQ-aware delay schedule, prioritizing structural simplicity, scalability, and a clean long-context operating point. The Global-Latent + Local Transformer model (\textbf{MOSS-TTS-Local-Transformer}) introduces an additional frame-local autoregressive module that is more complex but more learning-efficient, yielding stronger speaker preservation at smaller scale and a shorter time to first audio.

These components yield a practical speech generation foundation model with a broad capability set, including zero-shot voice cloning, token-level duration control, phoneme-/pinyin-level pronunciation control, multilingual synthesis with smooth code-switching (notably between Chinese and English), and stable long-form generation up to hour-scale outputs.

\paragraph{Contributions.}
This technical report makes the following contributions:
\begin{itemize}
  \item We present MOSS-TTS, a discrete-token autoregressive speech generation foundation model built on a scalable discrete + AR + pretraining recipe.
  \item We integrate and analyze \textbf{MOSS-Audio-Tokenizer} \citep{gong2026moss} as a universal, streaming-compatible audio tokenizer with variable bitrate and unified semantic-acoustic representations.
  \item We present a large-scale, high-quality data pipeline that supports training on millions of hours of data and enables robust multilingual pretraining and controllable synthesis behavior.
  \item We release and compare two complementary discrete AR architectures (Delay-Pattern vs.\ Global-Latent + Local Transformer) that expose a clear tradeoff between structural simplicity/scalability and modeling efficiency/quality.
  \item We demonstrate broad controllability features (voice cloning, duration control, pronunciation control) and strong empirical performance on speaker similarity and quality metrics.
\end{itemize}

\paragraph{Organization.}
The remainder of the report is organized as follows. We begin with an overview of related work. We then describe the audio tokenizer and the overall modeling architectures, followed by the pretraining data pipeline and training recipe. We next present the evaluation results before concluding.

\section{Related Work}

MOSS-TTS sits at the intersection of discrete audio tokenization, large-scale autoregressive sequence modeling, and speech generation foundation models. We review the most relevant directions below.

\paragraph{Neural audio codecs and discrete audio tokenization.}
Discrete representations have become a standard foundation for scalable audio generation, following the broader success of vector quantization in representation learning \citep{van2017neural}.
Neural codecs such as SoundStream \citep{zeghidour2021soundstream} and subsequent high-fidelity compression models \citep{defossez2022high,kumar2023high} demonstrate that a learned encoder--quantizer--decoder stack can support low-bitrate reconstruction while remaining compatible with downstream sequence modeling.
Recent toolkits and open implementations further accelerate codec research and adoption \citep{wu2023audiodec,du2024funcodec}.
For speech generation in particular, an effective tokenizer must not only reconstruct waveforms, but also expose tokens that are semantically aligned with text and robust under long-horizon generation.
Several recent lines explore semantic shortcomings and the semantic--acoustic tradeoff in codec tokens \citep{ye2025codec,defossez2024moshi,gong2025xy}, motivating tokenizers that better balance compression, perceptual quality, and text-aligned semantics.

\paragraph{Audio language modeling with discrete tokens.}
With discrete tokens, audio generation can be cast as token sequence modeling, enabling language-model-like scaling and training recipes \citep{borsos2023audiolm,wu2024towards,latif2023sparks}.
Codec language models have been shown to produce intelligible speech and even zero-shot TTS behavior when trained autoregressively over discrete units \citep{wang2023neural}.
Concurrently, a growing body of work studies how token choices and modeling decisions affect controllability, semantic fidelity, and efficiency \citep{zhang2024speechgpt,yang2025almtokenizer}.
MOSS-TTS follows this trend but emphasizes a tokenizer and modeling stack designed to scale end-to-end without external pretrained audio teachers, aligning the discrete token representation with the requirements of AR speech generation.

\paragraph{TTS architectures: AR, NAR, diffusion/flow, and foundation-model scaling.}
Classical neural TTS systems progressed from AR acoustic modeling and neural vocoders \citep{oord2016wavenet,wang2017tacotron,shen2018tacotron2} to faster and more controllable NAR frameworks \citep{ren2019fastspeech,ren2020fastspeech2}, flow-based and diffusion-based synthesis \citep{kim2020glowtts,popov2021gradtts}. End-to-end approaches such as VITS \citep{kim2021vits} further unified acoustic modeling and waveform generation, improving simplicity and sample quality. More recently, scaling-driven and token-centric systems increasingly combine discrete representations with AR backbones for robustness and controllability at scale \citep{betker2023better}, as reflected in recent open technical reports and large-scale systems such as Qwen3-TTS \citep{hu2026qwen3}, CosyVoice \citep{du2024cosyvoice}, CosyVoice 3 \citep{du2025cosyvoice}, Seed-TTS \citep{anastassiou2024seed}, Fish-Speech \citep{liao2024fish}, and FireRedTTS-2 \citep{xie2025fireredtts}. Across these efforts, a recurring theme is that scaling data and model capacity alone is insufficient without a well-chosen discrete tokenizer and a model design that remains compatible with streaming, controllability, and long-context stability. MOSS-TTS complements this line of work by focusing on a fully discrete tokenization pipeline and token modeling strategies that remain efficient for long-form synthesis, while explicitly comparing two autoregressive architectures under the same tokenizer and large-scale pretraining recipe.

\paragraph{Voice cloning and controllability.}
Practical TTS systems increasingly demand controllability beyond text content, including speaker identity (voice cloning), speaking rate/duration control, and fine-grained pronunciation control.
Zero-shot voice cloning and multilingual universal generation have been explored via large-scale generative models and conditioning mechanisms \citep{le2023voicebox,casanova2022yourtts,liu2023styletts2}.
Token-centric systems also enable control signals to be expressed directly in the discrete domain, which can simplify modeling and improve stability compared to waveform-level control.
MOSS-TTS emphasizes token-level duration control and phoneme-/pinyin-level pronunciation interfaces, aiming to make control explicit and composable.

\section{Audio Tokenizer}

\subsection{Motivation and Design Principles}
Audio tokenizers serve as the foundational bridge for native Audio Large Language Models (Audio LLMs), transforming continuous raw audio signals into discrete tokens that can be seamlessly processed within a unified generative framework. A unified audio tokenizer for speech LLMs must satisfy two primary requirements: enabling high-fidelity reconstruction of diverse audio signals and maintaining compatibility with the sequential nature of autoregressive modeling~\citep{defossez2024moshi,li2025baichuan,zhang2025mimo}. 

Existing approaches typically address these requirements through pretrained audio encoders (e.g., HuBERT, Whisper)~\citep{hsu2021hubert,radford2023robust,ye2025codec,li2025dualcodec}, multi-stage training pipelines~\citep{wu2023audiodec,welker2025flowdec}, or architecture-specific inductive biases such as specialized CNN structures~\citep{zeghidour2021soundstream,defossez2022high,kumar2023high}. These designs often introduce external dependencies and architectural constraints that hinder the seamless scaling of model capacity, data volume, and quantization levels. Drawing inspiration from the success of LLMs, where simple, scalable architectures trained on massive datasets have proven superior~\citep{kaplan2020scaling, henighan2020scaling}, we posit that the performance ceiling of audio tokenizers can be raised by adopting a similar philosophy. We advocate for a simple, end-to-end scalable architecture that minimizes reliance on external priors or complex heuristics, emphasizing joint optimization and large-scale data exposure.

To address these limitations and support high-quality speech synthesis in MOSS-TTS, we use \textbf{MOSS-Audio-Tokenizer}, a high-performance audio tokenizer based on the \textbf{CAT} (\textbf{C}ausal \textbf{A}udio \textbf{T}okenizer with \textbf{T}ransformer) architecture~\citep{gong2026moss}. MOSS-Audio-Tokenizer is characterized by the following core strengths:

\begin{itemize}
    \item \textbf{High Compression and Variable Bitrate:} The model achieves a significant compression ratio, converting 24kHz audio into a discrete representation at only 12.5 frames per second (fps). Utilizing a 32-layer Residual Vector Quantization (RVQ) mechanism, it supports flexible bitrate adjustment from 0.125 to 4 kbps, catering to various high-fidelity reconstruction requirements.
    \item \textbf{Pure Transformer Architecture:} Unlike traditional codecs that rely on complex, hand-crafted CNN or hybrid CNN-Transformer blocks, MOSS-Audio-Tokenizer adopts a minimalist causal Transformer-based design. This architecture is intentionally unencumbered by specialized inductive biases, making it remarkably simple to implement and highly efficient to scale up. With a substantial 1.6-billion-parameter capacity, the model demonstrates superior representation power, while its inherently causal nature ensures seamless, frame-level streaming inference.
    \item \textbf{Universal Audio Representation:} The model is pretrained on millions of hours of diverse audio data, including speech, music, and environmental sound effects, ensuring robust generalization across all audio domains.
    \item \textbf{Unified Semantic-Acoustic Modeling:} The discrete tokens produced by MOSS-Audio-Tokenizer preserve strong reconstruction quality while inherently capturing rich semantic information, making them ideally suited for autoregressive LLM modeling.
    \item \textbf{End-to-End Joint Optimization:} All components, including the encoder, quantizer, decoder, discriminators, and the LLM used for semantic alignment, are optimized jointly to maximize the model's performance ceiling.
\end{itemize}

\subsection{Architecture}
\begin{figure*}[t!]
  \centering
  \includegraphics[width=0.8\linewidth]{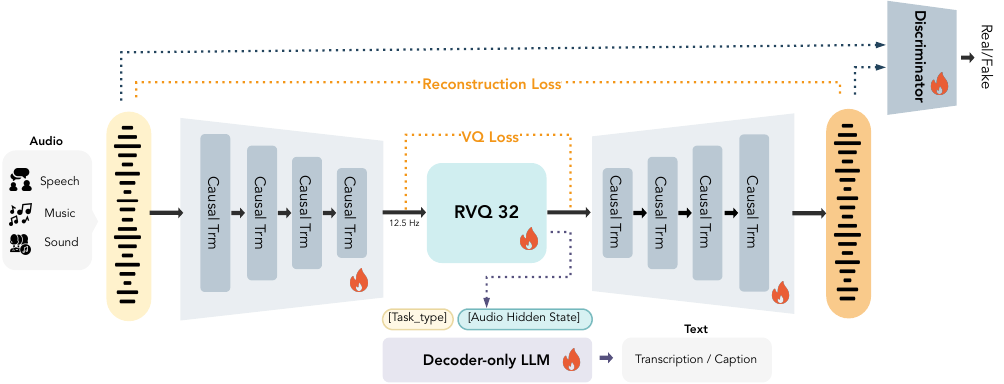}
  \caption{Architecture of MOSS-Audio-Tokenizer.
Both the encoder and decoder are built upon causal Transformers.
All components, including the encoder, quantizer, decoder, decoder-only LLM, and discriminator, are optimized jointly in an end-to-end manner.}
  \label{fig:audio_tokenizer_arch}
\end{figure*}

As illustrated in Figure~\ref{fig:audio_tokenizer_arch}, MOSS-Audio-Tokenizer adopts an RVQ-GAN framework for training. The model consists of five components: a causal encoder, a residual vector quantizer (RVQ), a causal decoder, a decoder-only LLM for semantic modeling, and adversarial discriminators.
\paragraph{Fully Transformer-based Encoder and Decoder.}
The encoder and decoder of MOSS-Audio-Tokenizer each consist of 68 causal Transformer blocks. To facilitate efficient streaming inference, both components use a 10-second sliding-window attention mechanism.
To progressively reduce the sequence length, the encoder incorporates patchify operations~\citep{dosovitskiy2020image} at the input stage and following layers 12, 24, and 36, with respective patch sizes of 240, 2, 2, and 2. Since these patchify operations alter the feature dimensionality, a linear projection is applied after each stage to map the hidden states to the corresponding Transformer block dimension. This configuration effectively downsamples raw 24\,kHz waveforms to a low frame rate of 12.5\,fps.
The encoder is structured into four stages with hidden dimensions of 768, 768, 768, and 1280, containing 12, 12, 12, and 32 Transformer blocks, respectively. For each stage, the feed-forward network (FFN) dimension is set to four times the hidden dimension. The multi-head self-attention mechanism uses 12, 12, 12, and 20 attention heads across the four stages. All Transformer blocks employ rotary positional embeddings (RoPE)~\citep{su2024roformer}. The decoder mirrors the encoder architecture in a fully causal manner. Both the encoder and decoder contain approximately 0.8B parameters and are trained from scratch.
\paragraph{Residual Vector Quantization.}
Discrete tokenization is performed using a 32-layer residual vector quantizer (RVQ). Each layer employs a codebook of size 1024 with factorized vector quantization (latent dimension 8)~\citep{kumar2023high} and L2-normalized codes. To enable variable-bitrate tokenization, random quantizer dropout~\citep{zeghidour2021soundstream} with a probability of 1.0 is applied during training.

\paragraph{Semantic Supervision.}
To encourage the learning of semantically structured discrete representations, we attach a 0.5B decoder-only causal language model~\citep{qwen2.5} as a semantic head. This head provides audio-to-text supervision by autoregressively predicting text conditioned on the quantizer outputs. The supervision tasks include Automatic Speech Recognition (ASR), multi-speaker ASR, and audio captioning.

\paragraph{Perceptual Modeling.}
To enhance the perceptual quality of the reconstructed audio, we employ a multi-period discriminator~\citep{defossez2022high} and a complex STFT discriminator~\citep{kumar2023high} for adversarial training with the audio tokenizer.

\subsection{Training}
MOSS-Audio-Tokenizer is trained on a massive dataset comprising millions of hours of both public and in-the-wild audio data. During training, we employ a multi-task learning framework to enable MOSS-Audio-Tokenizer to achieve both robust semantic alignment with text and high-fidelity audio reconstruction. The modeling approach for each component is detailed as follows.

\paragraph{Semantic Modeling via Audio-to-Text Tasks.}
To encourage the token representation to be semantically rich and aligned with text-based language modeling, we incorporate an auxiliary audio-to-text objective. Specifically, we employ a 0.5B-parameter decoder-only LLM~\citep{qwen2.5} and condition it on the representations produced by MOSS-Audio-Tokenizer. Concretely, we feed the hidden states from the quantizer output into the LLM, which then autoregressively predicts textual tokens. We consider a diverse set of audio-to-text tasks, including automatic speech recognition (ASR), multi-speaker ASR, and audio captioning. For audio samples that are paired with textual annotations, we apply the corresponding semantic modeling objective. Each task is specified by a fixed task tag \( \mathcal{T} \), which is prepended to the LLM input. The semantic objective is optimized using a standard cross-entropy loss:
\begin{equation}
\label{eq:audio_to_text_loss}
\mathcal{L}_{\mathrm{sem}}
=
- \sum_{t=1}^{|\mathrm{s}|}
\log p_{\theta_{\mathrm{LLM}}}
\left(
\mathrm{s}_t \,\middle|\,
\mathcal{T},\, \mathbf{q},\, \mathrm{s}_{<t}
\right),
\end{equation}
where $\mathbf{s} = (\mathrm{s}_1, \dots, \mathrm{s}_{|\mathbf{s}|})$
denotes the target text token sequence,
$\mathbf{q}$ denotes the sequence of quantized audio representations produced by  MOSS-Audio-Tokenizer,
$\mathcal{T}$ is a task-specific prompt token,
and $\theta_{\mathrm{LLM}}$ are the parameters of the causal language model.

\paragraph{Quantizer Optimization.}
For training simplicity and stability, each quantization layer in MOSS-Audio-Tokenizer adopts
factorized vector quantization~\citep{kumar2023high},
where codebooks are directly optimized via gradient descent,
without relying on additional codebook update mechanisms~\citep{defossez2022high}. We incorporate a commitment loss and a codebook loss to jointly
optimize the encoder and the codebook entries:
\begin{equation}
\label{eq:cmt_loss}
\mathcal{L}_{\mathrm{cmt}} = \sum_{c=1}^{N_q}
\left\| \mathbf{z}_c - \operatorname{sg}(q_c(\mathbf{z}_c)) \right\|_2^2,
\end{equation}
\begin{equation}
\label{eq:codebook_loss}
\mathcal{L}_{\mathrm{code}} = \sum_{c=1}^{N_q}
\left\| \operatorname{sg}(\mathbf{z}_c) - q_c(\mathbf{z}_c) \right\|_2^2,
\end{equation}
where \( \mathbf{z}_c \) denotes the input to the \( c \)-th quantization layer,
\( q_c(\mathbf{z}_c) \) is the corresponding quantized output,
\( N_q \) is the number of quantizers,
and \( \operatorname{sg}(\cdot) \) denotes the stop-gradient operator~\citep{van2017neural}.

\paragraph{Acoustic Modeling via Reconstruction Tasks.}
To ensure high-fidelity and domain-robust audio reconstruction, we adopt a multi-scale mel-spectrogram loss:
\begin{equation}
\label{eq:rec_loss}
\mathcal{L}_{\mathrm{rec}} =
\sum_{i=5}^{11}
\left\|
S_{2^i}(\mathbf{x}) - S_{2^i}(\hat{\mathbf{x}})
\right\|_1,
\end{equation}
where \( S_{2^i}(\cdot) \) denotes the mel-spectrogram computed using a normalized short-time Fourier transform (STFT) with window size \( 2^i \) and hop size \( 2^{i-2} \). Here, \( \mathbf{x} \) is the ground-truth waveform and \( \hat{\mathbf{x}} \) is the reconstructed waveform generated by the decoder.

\paragraph{Adversarial Training.}
To further improve reconstruction fidelity and perceptual quality, we employ adversarial training with multiple discriminators. The discriminator loss follows the least squares GAN (LSGAN) formulation~\citep{mao2017least}, given by:
\begin{equation}
\label{eq:disc_loss}
\mathcal{L}_\mathrm{D}(\mathbf{x}, \hat{\mathbf{x}}) = \frac{1}{K} \sum_{k=1}^{K}  (1 - D_k(\mathbf{x}))^2 + D_k^2(\hat{\mathbf{x}}),
\end{equation}
where \( D_k \) represents the \( k \)-th discriminator, \( K \) is the total number of discriminators, \( \mathbf{x} \) is the ground-truth audio, and \( \hat{\mathbf{x}} \) is the predicted audio.

For the generator, we include an adversarial loss and a feature matching loss. The adversarial loss encourages the generator to produce high-fidelity audio that is indistinguishable from real samples:
\begin{equation}
\label{eq:adv_loss}
\mathcal{L}_\mathrm{adv}(\hat{\mathbf{x}}) = \frac{1}{K} \sum_{k=1}^{K} (1 - D_k(\hat{\mathbf{x}}))^2.
\end{equation}
Additionally, we incorporate a feature matching loss $\mathcal{L}_{\mathrm{feat}}$~\citep{kumar2019melgan} to ensure structural similarity across multiple scales. It penalizes the $\ell_1$ distance between the intermediate feature maps of the discriminators for real and synthetic audio:
\begin{equation}
\label{eq:feature_matching_loss}
\mathcal{L}_\mathrm{feat}(\mathbf{x}, \hat{\mathbf{x}}) = \frac{1}{K} \sum_{k=1}^{K} \frac{1}{L_k} \sum_{l=1}^{L_k} \frac{ \left\| D_k^l(\mathbf{x}) - D_k^l(\hat{\mathbf{x}}) \right\|_1}{\operatorname{mean}(\left\| D_k^l( \mathbf{x}) \right\|_1)}
\end{equation}
where $D_k^l$ denotes the feature representation from the $l$-th layer of the $k$-th discriminator, and $L_k$ is the number of layers in that discriminator.

\paragraph{Overall Training Objective.}
The overall generator objective is a weighted combination of all loss terms:
\begin{equation}
\label{eq:generator_loss}
\begin{aligned}
\mathcal{L}_{\mathrm{G}} =\;&
\lambda_{\mathrm{sem}} \mathcal{L}_{\mathrm{sem}}
+ \lambda_{\mathrm{rec}} \mathcal{L}_{\mathrm{rec}}
+ \lambda_{\mathrm{cmt}} \mathcal{L}_{\mathrm{cmt}} 
+ \lambda_{\mathrm{code}} \mathcal{L}_{\mathrm{code}}
+ \lambda_{\mathrm{adv}} \mathcal{L}_{\mathrm{adv}}
+ \lambda_{\mathrm{feat}} \mathcal{L}_{\mathrm{feat}} ,
\end{aligned}
\end{equation}
where \( \lambda_{\mathrm{sem}} \), \( \lambda_{\mathrm{rec}} \), \( \lambda_{\mathrm{cmt}} \), \( \lambda_{\mathrm{code}} \), \( \lambda_{\mathrm{adv}} \), \( \lambda_{\mathrm{feat}} \) are scalar hyperparameters controlling the relative contribution of each loss term.


During training, we set the hyperparameters to $\lambda_{\mathrm{sem}}{=}20$, $\lambda_{\mathrm{rec}}{=}15$, $\lambda_{\mathrm{cmt}}{=}0.25$, $\lambda_{\mathrm{code}}{=}1.0$, $\lambda_{\mathrm{adv}}{=}1.0$,  $\lambda_{\mathrm{feat}}{=}2.0$.

Due to computational constraints, we adopt a two-stage training schedule to improve training efficiency: non-adversarial pretraining without discriminator-related losses for 520k steps (batch size 1536, approximately 5 hours of audio per batch), followed by adversarial fine-tuning for 500k steps (batch size 768). All modules are optimized end-to-end without pretrained encoders or semantic teachers~\citep{hsu2021hubert,radford2023robust,zhang2023speechtokenizer,defossez2024moshi,ye2025codec}.

\section{Architecture}

\begin{figure*}[t!]
  \centering
  \includegraphics[width=1.0\linewidth]{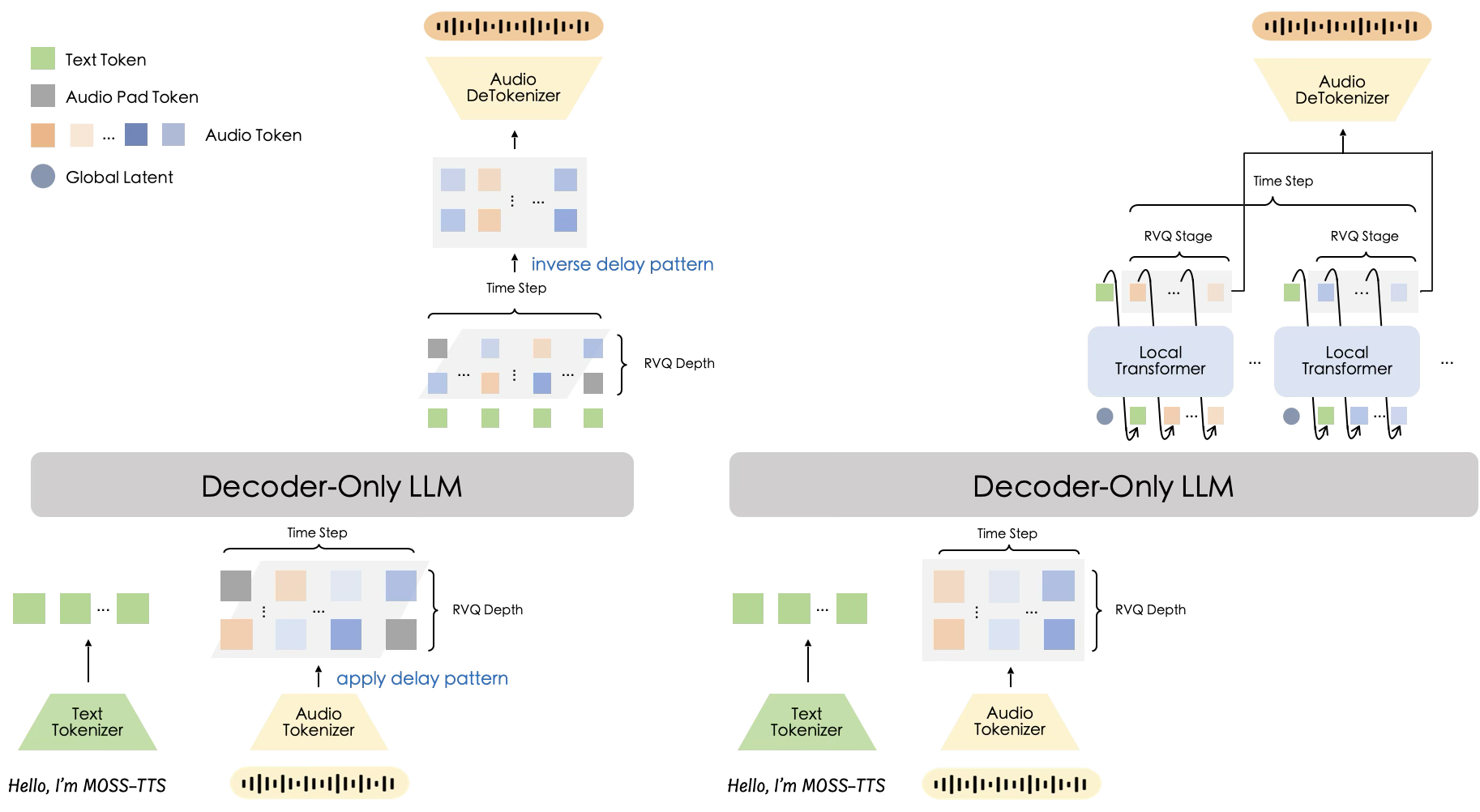}
  \caption{Architecture of MOSS-TTS. The left panel illustrates the \textbf{delay pattern} as described in Section~\ref{sec:delay_pattern}, while the right panel depicts the \textbf{local transformer pattern} as detailed in Section~\ref{sec:local_transformer_pattern}.}
  \label{fig:moss_tts_arch}
\end{figure*}

MOSS-TTS is a speech generation foundation model built upon discrete audio tokens. To facilitate effective scaling and capitalize on the success of large language models (LLMs), we adopt a straightforward end-to-end, purely autoregressive (AR) architecture. As illustrated in Figure~\ref{fig:moss_tts_arch}, given a text sequence and an optional speech prompt, MOSS-TTS generates the target token sequence through next-token prediction. The central architectural question is not whether to use AR modeling, but how to handle the multi-stream discrete token block produced by the tokenizer. For a 32-layer RVQ tokenizer, the chosen token modeling pattern directly determines engineering complexity, scaling behavior, decoding latency, and final synthesis quality.

Rather than committing to a single token modeling pattern a priori, we train two architectures under the same tokenizer and large-scale pretraining recipe. This serves a concrete research purpose: to isolate the effect of the token modeling pattern itself in large-scale discrete speech modeling. In practice, the two designs expose a clear tradeoff. \textbf{Delay Pattern} uses a structurally simple single-backbone, multi-head parameterization, making it easier to scale to large model sizes, long contexts, and optimized inference backends. \textbf{Local Transformer} introduces an additional frame-local autoregressive module, which increases architectural complexity but improves modeling efficiency; during internal development, it exhibited consistently lower per-layer token losses, and the later voice-cloning evaluations show stronger speaker similarity at much smaller model scale. Correspondingly, the current report uses \textbf{MOSS-TTS-Local-Transformer} to highlight the quality advantage of the local pattern on standard cloning benchmarks (Table~\ref{tab:seed_tts_eval_voice_cloning}), while \textbf{MOSS-TTS} serves as the main architecture for duration control, pronunciation control, and ultra-long generation (Tables~\ref{tab:duration_control}, \ref{tab:phoneme_control}, and \ref{tab:ultra_long_internal}).

The tokenizer emits \(N_q=32\) RVQ layers. In our implementation, both architectures predict \(N_h=N_q+1=33\) channels at each aligned step: one text-or-pad channel \(y_{0,t}\) and 32 audio channels \(y_{1,t}, \dots, y_{N_q,t}\), where \(y_{j,t}=\mathbf{a}_{j,t}\) for \(j \ge 1\). When step \(t\) corresponds to an audio frame, \(y_{0,t}\) is trained to emit a dedicated pad symbol; on text-only steps, it emits the normal text token. We use the same head-wise weighted cross-entropy in both architectures, with
\begin{equation}
\boldsymbol{\lambda} =
\begin{aligned}
(&1, 3, 3, 3, 2, 2, 2, 1, 1, 1, 1, 1, 1, 1, 1, 1, 1, \\
 &1, 1, 1, 1, 1, 1, 1, 1, 1, 1, 1, 1, 1, 1, 1, 1),
\end{aligned}
\end{equation}
which up-weights the earliest coarse RVQ layers while keeping unit weight on the text-or-pad channel and the remaining finer layers.

\subsection{Delay Pattern}
\label{sec:delay_pattern}
To model the RVQ hierarchy efficiently without increasing the sequence length to \(T \times N_q\), we adopt a \textbf{delay pattern} \citep{copet2023simple}. Among the two architectures we study, this is the simpler and more scalable design: a single Transformer backbone carries the full sequence model, and each prediction channel is obtained by a lightweight head projection from the backbone hidden state.

Let $\mathbf{s}$ denote the input text sequence and $\mathbf{A} \in \{1, \dots, V\}^{N_q \times T}$ be the audio token matrix, where $V$ denotes the codebook size of each RVQ layer, $N_q$ is the number of quantizers, and $T$ is the number of audio frames. Each element $\mathbf{a}_{j,t} \in \{1, \dots, V\}$ represents the token index at the $j$-th RVQ layer and $t$-th time frame. We apply a time-delay shift such that the $j$-th layer is shifted forward by $j-1$ frames. The delayed token matrix $\tilde{\mathbf{A}}$ is defined as:
\begin{equation}
\label{eq:delay_shift}
\tilde{\mathbf{a}}_{j, t} = \mathbf{a}_{j, t - (j-1)}, \quad t \in \{j, \dots, T+j-1\}.
\end{equation}

\paragraph{Input Embedding.}
On the input side of the backbone LLM, we use $N_q$ distinct speech embedding tables. For each time step $t$ in the delayed sequence, the input audio representation vector $\mathbf{h}_{t} \in \mathbb{R}^D$ is the sum of embeddings across all layers:
\begin{equation}
\label{eq:sum_embedding}
\mathbf{h}_{t} = \sum_{j=1}^{N_q} \operatorname{Emb}_j(\tilde{\mathbf{a}}_{j, t}),
\end{equation}
where $\operatorname{Emb}_j(\cdot)$ denotes the embedding lookup for the $j$-th codebook and $D$ is the model hidden dimension. Text tokens are embedded with the standard text embedding table; the delay mechanism applies only to the RVQ audio streams. The resulting vector sequence of length $T+N_q-1$ is concatenated with text embeddings as the backbone input.

\paragraph{Modeling Objective.}
On the output side, the hidden state $\mathbf{x}_t$ is passed through \(N_h=33\) heads: one text-or-pad head and 32 audio heads. Let \(\tilde{y}_{0,t}=y_{0,t}\) and \(\tilde{y}_{j,t}={\mathbf{a}}_{j,t}\) for \(j \ge 1\). The weighted training objective is
\begin{equation}
    \mathcal{L}_{\mathrm{delay}} = -\sum_{t=1}^{T+N_q-1} \sum_{j=0}^{N_q} \lambda_j m_{j,t}\log p_{\theta_{\mathrm{delay}}}(\tilde{y}_{j,t} \mid \mathbf{E}, \{\tilde{y}_{x, y} : x + y < j + t + \mathrm{max}(0,\,1-j)\}),
\end{equation}
where \(\theta_{\mathrm{delay}}\) encompasses the parameters of the backbone, embeddings, and prediction heads; \(\mathbf{E}\) represents the text token sequence; and \(m_{j,t}\) masks the invalid positions introduced by delay shifting and padding. Because every channel is predicted directly from the backbone state, the delay pattern keeps the decoding path simple: once \(\mathbf{x}_t\) is available, token generation only requires head projections. This simplicity is one of the main reasons it is easier to implement, scale, and deploy.

\subsection{Local Transformer}
\label{sec:local_transformer_pattern}
We further explore a hierarchical token modeling design using a \textbf{Local Transformer}, inspired by the RQ-Transformer in Moshi~\citep{defossez2024moshi}. Unlike the delay pattern, this approach models the token block without introducing temporal shifts: the backbone produces one global latent per aligned step, and a lightweight autoregressive module expands that latent into the within-step token block. This design is architecturally more complex, but it offers a stronger inductive bias for frame-level token modeling.

\paragraph{Input Embedding.}
On the input side, we directly sum the embeddings of all RVQ layers at each time step $t$ without any delay. The input hidden state $\mathbf{h}_{t}$ to the backbone LLM is given by:
\begin{equation}
\label{eq:local_sum_embedding}
\mathbf{h}_{t} = \sum_{j=1}^{N_q} \operatorname{Emb}_j(\mathbf{a}_{j, t}),
\end{equation}
where $\mathbf{a}_{j, t}$ is the token at the $j$-th RVQ layer and $t$-th time frame. 
\paragraph{Hierarchical Decoding.}
On the output side, we employ a lightweight \textbf{Local Transformer} to autoregressively decode the full per-step token block. Specifically, let \(\mathbf{x}_t\) be the output hidden state from the backbone LLM at time \(t\). The Local Transformer predicts the sequence \((y_{0,t+1}, y_{1,t+1}, \dots, y_{N_q,t+1})\) sequentially. The input to the Local Transformer when predicting channel \(j\), denoted as \(\mathbf{z}_{j, t}\), is defined as:
\begin{equation}
\label{eq:local_input}
\mathbf{z}_{j, t} = 
\begin{cases} 
\mathbf{x}_t, & \text{if } j = 0, \\
\operatorname{Emb}^{\mathrm{}}_{j-1}(y_{j-1, t+1}), & \text{if } 1 \le j \le N_q.
\end{cases}
\end{equation}
The Local Transformer processes \(\mathbf{z}_{j, t}\) and passes the resulting hidden states through the corresponding prediction head to emit \(y_{j,t+1}\).

\paragraph{Modeling Objective.}
The entire architecture, including the backbone and the local transformer, is trained end-to-end and optimized via
\begin{equation}
\label{eq:local_loss}
\mathcal{L}_{\mathrm{local}} = - \sum_{t=1}^{T} \sum_{j=0}^{N_q} \lambda_j \log p_{\theta_{\mathrm{local}}}(y_{j, t} \mid \mathbf{E}, y_{<j, t}, y_{:, <t}),
\end{equation}
where \(y_{<j, t}\) denotes the preceding channels at the current aligned step, and \(y_{:, <t}\) denotes all channels from previous steps. \(\theta_{\mathrm{local}}\) encompasses the parameters of the backbone LLM, the local transformer, embeddings, and prediction heads. Compared with the delay pattern, this design inserts an additional autoregressive loop of length \(N_q+1\) inside each frame. As a result, it is computationally heavier in steady-state decoding, but it can start emitting audio earlier because it does not need to wait for delayed offsets to materialize the first frame. Empirically, its main advantage in this report is not architectural simplicity, but higher modeling efficiency and stronger speaker preservation.

Moreover, we incorporate \textit{Progressive Sequence Dropout} as proposed in MOSS-Audio-Tokenizer~\citep{gong2026moss} to support bitrate-controllable audio generation.

\section{Pretraining}

\subsection{Pretraining Data}
\label{subsec:pretraining_data}
\begin{figure}[t]
  \centering
  \includegraphics[width=\linewidth]{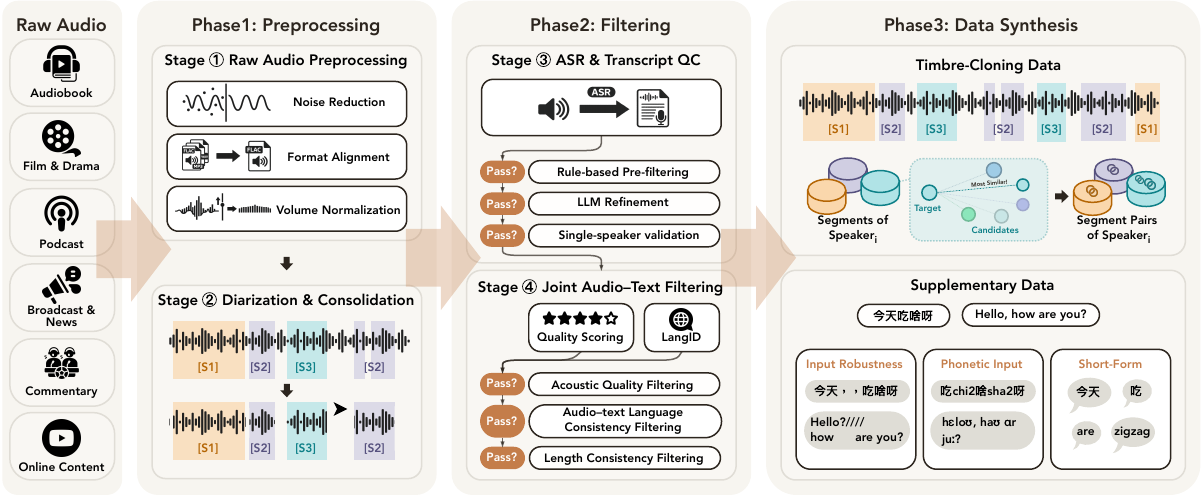}
  \caption{Overview of the MOSS-TTS pretraining data pipeline, including preprocessing, filtering, and targeted data synthesis.}
  \label{fig:pretraining_data_overview}
\end{figure}

Scaling TTS pretraining to millions of hours of speech data necessitates sourcing audio from naturally occurring, open-domain recordings---podcasts, audiobooks, broadcast \& news, film \& drama, commentary, and online content. Such recordings, however, rarely satisfy the conditions required for direct TTS supervision: they routinely contain multiple concurrent speakers, background music, ambient noise, and unreliable or missing transcription metadata. High-quality pretraining therefore demands that two fundamental properties hold for every training unit: (i)~the audio is acoustically clean and contains the speech of a single speaker, free from overlapping voices, music, and significant background noise, and (ii)~the paired transcript is linguistically well-formed and faithfully aligned to the spoken content. To enforce these properties at scale, we design a multi-stage data pipeline that progressively transforms raw web audio into curated, trainable speech--transcript pairs. As summarized in Figure~\ref{fig:pretraining_data_overview}, the pipeline is organized into three phases. The \textbf{preprocessing} phase (Stages~1--2) establishes a standardized acoustic foundation and extracts speaker-consistent segments via diarization. The \textbf{filtering} phase (Stages~3--4) first produces and refines transcripts with multilingual ASR, rule-based checks, and LLM-based quality control, and then retains only pairs that pass joint audio--transcript filtering, including acoustic quality checks, audio/text language-consistency checks, and duration-text consistency checks. The \textbf{data synthesis} phase supplements the corpus with targeted examples that introduce explicit speaker-conditioning structure for timbre transfer, broaden coverage of underrepresented input types, and strengthen robustness to diverse real-world input formats.

\subsubsection{Data Preprocessing}
\label{subsubsec:pretrain_preprocessing}
As shown in Figure~\ref{fig:pretraining_data_overview}, the preprocessing phase contains Stages~1--2 and prepares acoustically standardized, speaker-consistent segments before any transcript is produced.

\textbf{Stage 1: Raw audio preprocessing.}
Raw web-sourced recordings exhibit substantial heterogeneity in acoustic format: sampling rates vary widely across sources, loudness levels differ by tens of decibels, and many files carry environmental noise, music accompaniment, or reverberation. Left uncorrected, this variability degrades the reliability of both speaker diarization (Stage~2) and ASR (Stage~3), and introduces inconsistency into the acoustic features used during model training. We therefore apply the following preprocessing steps to each recording as the first pipeline stage:
\begin{itemize}
  \item \textbf{Noise suppression.} We apply MossFormer2-SE-48K \citep{zhao2024mossformer2}, a neural speech enhancement model, to suppress stationary and non-stationary background noise. The purpose of denoising at this stage is not to produce the final training signal but to improve the reliability of downstream speaker diarization: a cleaner input yields more accurate voice activity detection and sharper speaker boundary estimates. Audio is resampled to 48\,kHz before enhancement, which is the native operating rate of the model.

  \item \textbf{Format standardization.} Parameter alignment such as sample type, channel layout, and header metadata is enforced after enhancement, and the processed output is written to FLAC format, establishing a uniform format contract for all subsequent stages.

  \item \textbf{Volume normalization.} To reduce inter-source level variation, we apply a two-stage gain procedure. First, we compute the RMS-based signal level in dBFS,
  \[
    L_{\mathrm{dBFS}}(\mathbf{x}) = 20\log_{10}\!\left(\sqrt{\tfrac{1}{T}\textstyle\sum_{t}x_t^2} + \epsilon\right),
  \]
  and apply a clamped gain $g = \operatorname{clip}(-20 - L_{\mathrm{dBFS}}(\mathbf{x}),\; -3,\; 3)$\,dB, rescaling the waveform by $10^{g/20}$. The target of $-20$\,dBFS and the $\pm 3$\,dB clipping range together prevent both over-compression of already-quiet recordings and excessive amplification of outliers. Second, peak normalization divides by the maximum absolute sample value to map the result to the $[-1, 1]$ amplitude range, ensuring numerical consistency across the pipeline.
\end{itemize}

\textbf{Stage 2: Speaker diarization and segment consolidation.}
We run speaker diarization on the denoised audio to obtain a time-ordered sequence of speaker-labeled intervals,
\[
  \mathcal{D} = \{(k_i,\, t_i^{\mathrm{st}},\, t_i^{\mathrm{ed}})\}_{i=1}^{N},
\]
where $k_i$ is a recording-local speaker label (e.g., \texttt{SPEAKER-00}, \texttt{SPEAKER-01}, \dots) and $t_i^{\mathrm{st}} < t_i^{\mathrm{ed}}$ are the start and end timestamps. Speaker labels are meaningful only within a single recording; we do not perform cross-recording identity linking. We use DiariZen \citep{han2025efficient,han2025fine,han2025leveraging}, an end-to-end neural diarization system, for this step.

Raw diarization output is typically fragmented: a single continuous speaking turn may be split into multiple short intervals separated by brief pauses or breath sounds. Training directly on such fragments would over-represent short, sub-sentence units at the expense of longer, paragraph-level continuity. We therefore apply a two-step consolidation procedure to maximize contiguous single-speaker coverage:
\begin{itemize}
  \item \textbf{Filtering and consecutive-speaker merging.} Segments shorter than $\tau_{\min} = 0.1$\,s are discarded as unreliable diarization artifacts. The remaining segments are then scanned in chronological order: whenever two adjacent segments carry the same speaker label, they are merged into a single interval spanning both. This produces a consolidated sequence
  \[
    \mathcal{A} = \{(k_j,\, s_j,\, e_j)\}_{j=1}^{M}, \quad M \le N,
  \]
  where each run of consecutive same-speaker segments in $\mathcal{D}$ has been collapsed into one entry. There is no gap threshold on merging: any two adjacent same-speaker segments are unified regardless of the intervening silence duration, because such gaps reflect natural pauses within a speaking turn rather than speaker changes.

  \item \textbf{Single-speaker truncation.} We apply a hard one-hour limit to avoid unbounded unit lengths. Starting from $s_1$ (the onset of $\mathcal{A}$), we define a cutoff $t_{\mathrm{lim}} = s_1 + 3600$\,s and emit segments from $\mathcal{A}$ in order, clamping the endpoint of the last included segment to $t_{\mathrm{lim}}$ and discarding any resulting segment shorter than $\tau_{\min}$. The output is a list of consolidated, speaker-labeled intervals drawn from at most one hour of the recording, starting from the onset of the first diarized segment.
\end{itemize}
\subsubsection{Data Filtering}
\label{subsubsec:pretrain_filtering}
As shown in Figure~\ref{fig:pretraining_data_overview}, the filtering phase corresponds to Stages~3--4: Stage~3 constructs and cleans transcripts, and Stage~4 keeps only pairs that are jointly consistent on the audio and transcript sides.

\textbf{Stage 3: ASR and transcript quality control.}
Each consolidated segment from Stage~2 passes through a sequential pipeline that transcribes the audio and then applies a series of quality-control steps to produce a clean, speaker-tagged transcript suitable for TTS training.

\begin{itemize}
  \item \textbf{ASR transcription.} We transcribe each segment using MOSS-Transcribe-Diarize \citep{yu2026moss}, our proprietary multilingual ASR model. The model does not require an externally provided language label; it directly produces a multilingual diarization-aware transcript. The raw output follows a structured format in which every utterance span is prefixed by a recording-local speaker tag (e.g., \texttt{[S1]}, \texttt{[S2]}) and may contain inline sound-event markers (e.g., \texttt{[music]}, \texttt{[laugh]}). This structured output reflects the full diarization-aware recognition result and requires subsequent cleaning before use as a training transcript.

  \item \textbf{Rule-based pre-filtering.} Before invoking the LLM, three lightweight rules discard clearly unusable transcripts to avoid wasting inference budget:
  \begin{itemize}
    \item \textit{Empty content}: the transcript is blank or consists only of whitespace after stripping.
    \item \textit{Severe repetition loop}: any phrase is repeated consecutively more than six times, which is a reliable signal of ASR model collapse.
    \item \textit{Non-speech dominance}: after removing all bracketed tags (\texttt{[...]}), the remaining linguistic content accounts for less than 20\% of the total text length, indicating that the segment is predominantly noise, music, or other non-speech events.
  \end{itemize}
  Segments failing any rule are discarded without further processing.

  \item \textbf{LLM-based transcript refinement.} Segments that pass pre-filtering are processed by a large language model using a structured two-stage prompt.
  \begin{itemize}
    \item \textit{Diagnosis (filtering)}: the LLM first checks for two fatal defects. \texttt{filter-1} targets identical spoken content repeated two or more times (distinct from the trivially repeating speaker tags that are a normal output of MOSS-Transcribe-Diarize). \texttt{filter-2} targets sentence truncation, identified by a trailing hyphen or an abrupt mid-word termination. Segments receiving either code are discarded.
    \item \textit{Correction (refinement)}: segments that pass diagnosis undergo sequential cleaning. \texttt{refine-1} removes all non-speech event tags while preserving speaker tags and linguistic content. \texttt{refine-2} deletes any speaker tag that is left with no following speech content. \texttt{refine-3} applies minimal structural repair to restore the standard \texttt{[speaker]content} format without modifying the recognized words. Not all steps are applied to every segment; segments already in correct form receive a \texttt{no-change} code and are passed through unmodified.
  \end{itemize}
  If the LLM call fails (e.g., due to a malformed response), the segment is discarded rather than falling back to the uncleaned transcript.

  \item \textbf{Single-speaker transcript validation.} As a final check, we verify that the refined transcript contains only \texttt{[S1]} speaker tags. The presence of any \texttt{[S2]}, \texttt{[S3]}, or higher-indexed tag indicates that multiple speakers were detected within the segment at the transcription level, which is inconsistent with the single-speaker constraint enforced in Stage~2. Such segments are discarded.
\end{itemize}

\textbf{Stage 4: Joint audio--transcript filtering.}
Segments that survive Stage~3 undergo a second round of filtering that combines acoustic quality signals with audio--transcript consistency checks. Unlike Stage~3, which focuses on transcript quality, this stage treats the audio and transcript jointly; a segment is retained only if both its acoustic quality and its audio--transcript consistency fall within acceptable bounds.

\begin{itemize}
  \item \textbf{Acoustic quality filtering.} We compute DNSMOS \citep{reddy2022dnsmos} and Meta AudioBox Production Quality (PQ) \citep{tjandra2025meta} on the pre-denoising audio rather than on the enhanced signal, because speech enhancement can distort quality estimates and make the scores reflect the enhancer rather than the source recording. A segment is accepted only if its DNSMOS score exceeds $2.8$ and its Meta AudioBox PQ score exceeds $6.5$, retaining only segments with sufficiently clean and natural-sounding speech.

  \item \textbf{Audio--text language consistency filtering.} We derive two language labels from different modalities and require them to agree. First, Whisper large-v3 \citep{radford2023robust} is applied to the audio to obtain an audio-side language label \(\hat{\ell}^{\mathrm{aud}}\). Second, a large language model reads the refined transcript and predicts a text-side language label \(\hat{\ell}^{\mathrm{text}}\). We retain a segment only if \(\hat{\ell}^{\mathrm{aud}} = \hat{\ell}^{\mathrm{text}}\). This removes pairs whose transcript language is inconsistent with the spoken content or whose ASR output is unreliable enough to confuse transcript-side language identification. For the remaining segments, we denote the agreed label by \(\hat{\ell}\).

  \item \textbf{Audio--transcript length consistency filtering.} A systematic mismatch between audio duration and transcript length indicates one of two failure modes: (i)~the audio is far longer than the transcript, suggesting that large portions of the segment are silence or non-speech background; or (ii)~the transcript is far longer than the audio, a reliable indicator of ASR hallucination. To detect both cases, we compute a language-specific character rate,
  \[
    r = \frac{|x'|}{d},
  \]
  where $|x'|$ is the character count of the refined transcript and $d$ is the segment duration in seconds. For each supported language $\ell$, we define a valid rate interval $[r_\ell^{\min},\, r_\ell^{\max}]$ derived from empirical statistics over a reference corpus; the agreed language label \(\hat{\ell}\) from the previous step is used to select the appropriate bounds. Segments whose rate $r \notin [r_\ell^{\min},\, r_\ell^{\max}]$ are discarded.
\end{itemize}
\subsubsection{Data Synthesis}
\label{subsubsec:pretrain_synthesis}
As shown in Figure~\ref{fig:pretraining_data_overview}, the final phase augments the naturally filtered corpus with targeted synthetic or transformed examples that cover capabilities not directly available from organic web audio.

Even after the pipeline described above, the filtered web-sourced corpus still leaves three systematic gaps that cannot be closed by filtering alone. The most consequential is the absence of explicit speaker-conditioning structure: the filtered corpus pairs text with speech but provides no prompt audio, and a model trained on it alone has no mechanism to transfer timbre from a reference speaker. The remaining two gaps are on the text side: real user inputs often contain formatting noise that the model must handle gracefully , for example, inputs such as ``Hello??!! are you there'' or ``Iwant to knowwhere it\quad is'', and phonetic script input is absent from organic speech data yet required for fine-grained pronunciation control. We address all three through targeted data synthesis.

\paragraph{Timbre-cloning data construction.}
The goal of this construction is to provide (prompt audio, target audio) pairs in which both sides originate from the same speaker, enabling the model to learn prompt-conditioned timbre transfer from real recorded speech rather than from any generative process. Construction proceeds entirely from the filtered corpus produced by Stages~1--4.

For each recording, we group the surviving segments by their diarization-assigned speaker identity. Let the segments attributed to a given speaker be $\{s_1, s_2, \dots, s_n\}$. For each target segment $s_i$, we construct a prompt candidate pool as follows. For every other segment $s_j$ ($j \neq i$), we draw five random temporal crops of $s_j$, each with independently sampled start and end timestamps subject to a maximum duration of 30\,s, yielding $5(n-1)$ prompt candidates in total. We then score every candidate against $s_i$ by computing the cosine similarity between their speaker embeddings, extracted using the fine-tuned WavLM-Large model employed in Seed-TTS-eval \citep{anastassiou2024seed}. The candidate attaining the highest similarity is selected as the prompt for $s_i$, producing the final training pair $(\text{prompt} = s_{j,\text{partial}}^{*},\; \text{target} = s_i)$.

This construction strategy has two important properties. First, by evaluating similarity on the cropped partials rather than on full segments, the selection directly optimizes for how well the prompt conveys the speaker's identity at inference-time durations, rather than for full-segment representativeness. Second, restricting the prompt to at most 30\,s and randomizing its boundaries encourages the model to extract stable timbre representations from varied temporal windows, improving robustness to prompt length and position at inference time.

\paragraph{Supplementary data.}
Three smaller supplements address remaining distributional gaps. For \textbf{input robustness}, we apply four noise transformations to the text of existing validated pairs---punctuation noise (consecutive marks, mixed full/half-width forms, malformed combinations), whitespace artifacts (extra spaces, misplaced line breaks), punctuation dropout, and sparse dirty-character injection---without modifying the audio. For \textbf{phonetic script input}, we support both full-sentence and partial (word- or phrase-level) replacement of orthographic text with phonetic notation: tone-marked pinyin for Chinese (e.g., \texttt{nin2 hao3}) and IPA enclosed in slashes for English (e.g., \texttt{/hæŋ bæk/}); training pairs are derived from filtered corpus segments by rule-based transcript conversion, with audio left unchanged. For \textbf{dictionary-style short-form data}, we supplement the corpus with single-character and single-word utterances, which are severely underrepresented in web-sourced recordings yet constitute a common real-world input pattern; without targeted coverage, a model trained on organic data alone tends to be unreliable on such ultra-short inputs.

\begin{figure}[t]
  \centering
  \includegraphics[width=\linewidth]{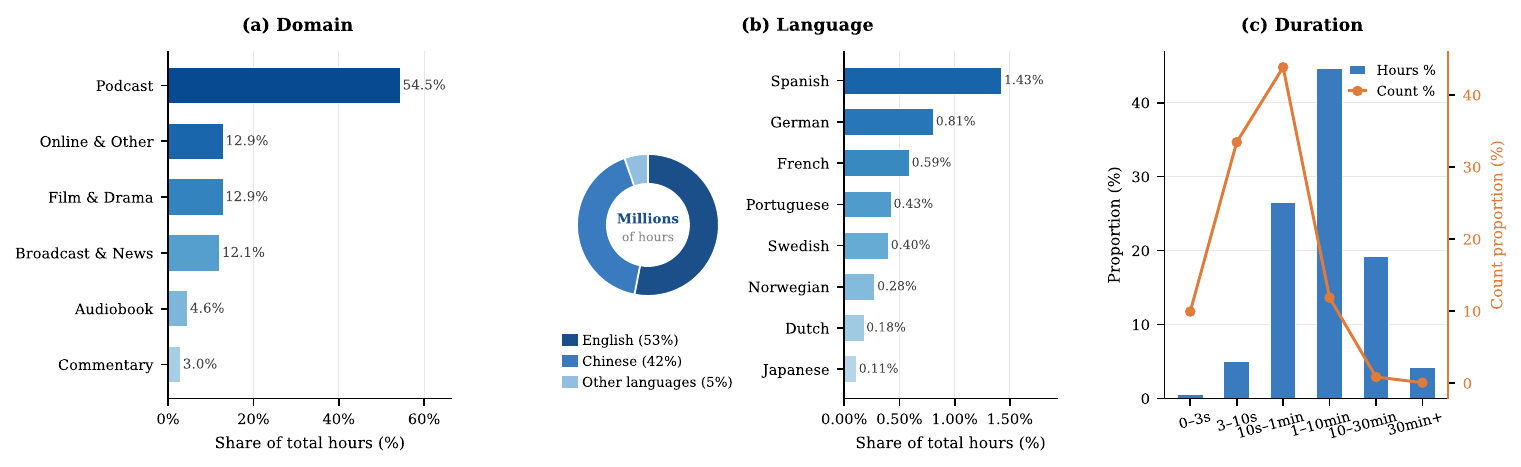}
  \caption{Statistics of the MOSS-TTS pretraining corpus. Panel~(a) shows the share of training hours by domain; panel~(b) shows the language distribution as a donut chart (English/Chinese/Other) alongside a breakdown of the top minor languages; panel~(c) shows the distribution of utterance duration by both hours (bars) and utterance count (line).}
  \label{fig:corpus_statistics}
\end{figure}

Figure~\ref{fig:corpus_statistics} summarizes the composition of the resulting corpus across all data subsets, showing the distribution of training hours by domain, language, and utterance duration.

For duration control, every training asset is serialized into two parallel variants. In the duration-conditioned variant, the prompt field \texttt{tokens} stores an explicit integer equal to the target audio-token count; in the free-duration variant, the same field is set to \texttt{None}. This paired formatting is applied uniformly across the corpus, so explicit and implicit duration supervision are both present throughout pretraining rather than being introduced only in a specific phase.

\subsection{Pretraining Stage}
\label{subsec:pretraining_stage}
For brevity, we denote the five training data subsets used in the curriculum as follows: the main filtered corpus \(\mathcal{D}_{\mathrm{basic}}\), timbre-cloning pairs \(\mathcal{D}_{\mathrm{clone}}\), dictionary-style short-form data \(\mathcal{D}_{\mathrm{dict}}\), noisy-text augmentation data \(\mathcal{D}_{\mathrm{noise}}\), and phonetic augmentation data \(\mathcal{D}_{\mathrm{phone}}\). The curriculum design follows three principles. First, we begin with the highest-density and least ambiguous supervision to maximize early learning efficiency. Second, harder conditioning tasks such as voice cloning and pronunciation control are introduced while the optimizer is still in a high-learning-rate stable region, so that they become native behaviors rather than narrow later-stage patches. Third, long-context extension is deferred until the short-context model has largely converged, which substantially reduces optimization instability and preserves short-utterance quality. The full schedule follows a simple warmup--stable--decay (WSD) pattern \citep{hu2024minicpm}: the learning rate is warmed up only in Phase~1, held fixed in the other non-decay phases, and linearly decayed from \(2\times10^{-4}\) to \(2\times10^{-6}\) only in Phase~3.

\begin{table}[t]
\centering
\caption{\textbf{Four-phase pretraining schedule of MOSS-TTS.}}
\label{tab:pretraining_curriculum}
\small
\setlength{\tabcolsep}{4pt}
\renewcommand{\arraystretch}{1.08}
\begin{tabular}{@{} >{\centering\arraybackslash}m{0.11\linewidth} >{\centering\arraybackslash}m{0.14\linewidth} >{\raggedright\arraybackslash}m{0.66\linewidth} @{}}
\toprule
\textbf{Phase} & \textbf{Max seq.} & \textbf{LR Schedule and Data Mixture Updates} \\
\midrule
P1 & 32k & LR warmup to \(2\times10^{-4}\), then hold; use \(\mathcal{D}_{\mathrm{basic}}\) only. \\
P2 & 32k & Hold LR at \(2\times10^{-4}\); enable all data subsets and strongly upsample \(\mathcal{D}_{\mathrm{clone}}\). \\
P3 & 32k & Linearly decay LR from \(2\times10^{-4}\) to \(2\times10^{-6}\); keep all subsets active and restore the data mixture to normal proportions. \\
P4 & 64k & Hold LR at \(2\times10^{-6}\); retain the rebalanced full-data mixture, expand the context window from 32k to 64k, and heavily upsample long-form data. \\
\bottomrule
\end{tabular}
\end{table}

\paragraph{Phase 1: basic alignment acquisition.}
We begin with \(\mathcal{D}_{\mathrm{basic}}\) only, covering Chinese, English, and lower-resource languages under relatively clean and direct text--speech supervision. This phase includes the only explicit warmup in the whole training schedule: the learning rate is first increased to \(2\times10^{-4}\) and then held fixed for the remainder of the phase. Excluding more specialized objectives at this stage improves sample efficiency: the model first learns monotonic text-to-audio alignment, multilingual grapheme-to-acoustic mapping, and the basic semantics encoded by the tokenizer before it is asked to solve voice transfer or pronunciation-correction tasks. Empirically, this stage produces a substantially stronger initialization for the subsequent mixed-data phases than training on the full heterogeneous mixture from step zero.

\paragraph{Phase 2: capability expansion under stable high LR.}
After the base mapping is established, we switch to the full data universe and deliberately assign a much higher sampling weight to \(\mathcal{D}_{\mathrm{clone}}\). The reason is strategic: prompt-conditioned timbre transfer is both harder and more fragile than ordinary text-to-speech, and if it is introduced too weakly it tends to remain a tail capability. Keeping the learning rate fixed at \(2\times10^{-4}\) while the model is exposed to the full control-oriented data mixture allows the backbone to absorb cloning, dictionary reading, noisy-text robustness, and phonetic prompting as first-class behaviors rather than as late patches.

\paragraph{Phase 3: linear-decay mixture rebalancing and quality consolidation.}
Once the control capabilities are in place, we keep the full data universe active but restore the mixture to its normal proportions, while linearly decaying the learning rate from \(2\times10^{-4}\) to \(2\times10^{-6}\) over the entire phase. This step is critical. Oversampling timbre-cloning data for too long biases the model toward prompt copying and can suppress the relative influence of standard multilingual TTS, dictionary coverage, and robustness-oriented augmentations. Phase~3 therefore serves as the main quality-consolidation stage: the model revisits the full task distribution while the optimizer transitions from a still-flexible high-LR regime to a highly conservative low-LR regime. In WSD terms, this is the decay segment in which most of the final gains are consolidated \citep{hu2024minicpm}. Early in the phase, the remaining relatively large updates are still sufficient to repair mixture imbalance and absorb residual capability gaps; late in the phase, the much smaller updates improve stability, reduce hallucination-like failures, and sharpen the final tradeoff among intelligibility, speaker similarity, and controllability.

\paragraph{Phase 4: long-context extension.}
In the final phase, we keep the learning rate fixed at \(2\times10^{-6}\), increase the maximum sequence length from 32k to 64k, and heavily upsample long-form data. We intentionally do not introduce this longer context earlier. Training with a very long window from the beginning is significantly less efficient, because most examples do not require it and because early optimization is better spent learning the core text--speech mapping than fitting long-range attention patterns. Instead, we follow a late context-extension strategy analogous to recent LLM and TTS systems: once the base distribution has converged at moderate context length, the model is adapted to longer contexts under a small learning rate, which preserves short-form quality while teaching paragraph-scale and hour-scale continuity \citep{dubey2024llama,yang2025qwen3,hu2026qwen3}. The heavy upsampling of long-form data is important here: without it, the nominal 64k window would be underutilized by the natural length distribution of the corpus. This stage is intended to improve speaker consistency across long generations, reduce drift in prosody and content over extended passages, and enable the model to use longer prompt speech without destabilizing decoding.

\paragraph{Learning-rate shape and practical rationale.}
Viewed globally, the four phases form a simple WSD-style training program rather than four disconnected runs: a short warmup embedded in Phase~1 lifts the learning rate to \(2\times10^{-4}\), Phases~1--2 then share a stable plateau at \(2\times10^{-4}\), Phase~3 linearly decays the learning rate from \(2\times10^{-4}\) to \(2\times10^{-6}\), and Phase~4 holds the final low learning rate at \(2\times10^{-6}\) during long-context adaptation. This schedule combines the optimization efficiency of a long stable high-LR region with the reliability of a gradual decay into a low-LR refinement regime. The distinction matters in practice: the stable plateau is where the model acquires its main multilingual TTS and controllability behaviors, whereas the linear decay phase is where those behaviors are rebalanced and polished without the abrupt optimization shock that would come from dropping directly from \(2\times10^{-4}\) to \(2\times10^{-6}\). By the time training enters Phase~4, the optimizer is already in a conservative regime, which allows us to upsample long-form data and extend the context window with minimal damage to established short-form quality. Compared with a one-shot full-data recipe, the staged curriculum yields a better division of labor across phases: P1 learns the multilingual TTS prior, P2 makes control abilities robust, P3 restores distributional balance while progressively refining the model, and P4 transfers the already-competent model to longer contexts. This progression serves as a practical compromise between training efficiency, controllability, and long-form robustness, and it forms the default pretraining recipe used for the full MOSS-TTS release.

\section{Evaluation}

We evaluate MOSS-TTS from two complementary perspectives: (i) the audio tokenizer---whether it provides high-fidelity and semantically usable units across bitrates and domains---and (ii) the speech generation model---whether discrete autoregressive modeling and large-scale pretraining yield strong zero-shot voice cloning, multilingual robustness, token-level duration control, phoneme-/pinyin-level pronunciation control, and ultra-long speech generation. For the speech generation model, we report results for both \textbf{MOSS-TTS} and \textbf{MOSS-TTS-Local-Transformer}. Following influential TTS technical reports \citep{hu2026qwen3,du2025cosyvoice,zhou2025voxcpm}, we prioritize objective metrics that are easy to reproduce and interpret: content consistency measured by WER/CER using a fixed ASR backend, speaker similarity (SIM) measured by cosine similarity of pretrained speaker embeddings, and task-specific metrics for controllability and long-form behavior.

\subsection{Audio Tokenizer}

\begin{table*}[t]
  \centering
  \fontsize{8pt}{10pt}\selectfont
  \setlength{\tabcolsep}{4pt}
  \renewcommand{\arraystretch}{1.15}
  \caption{Reconstruction quality comparison of open-source audio tokenizers on speech and audio/music data.
Speech metrics are evaluated on LibriSpeech test-clean (English) and AISHELL-2 (Chinese) and reported as English/Chinese.
Audio metrics are evaluated on the AudioSet evaluation subset, while music metrics are evaluated on the MUSDB dataset; values are reported as audio/music.
STFT-Dist. denotes the STFT distance.
Higher is better for speech metrics, whereas lower is better for audio/music metrics. $\boldsymbol{N}_{\mathrm{VQ}}$ denotes the number of quantizers. \textbf{Bold} entries indicate the best result within each bitrate regime.}
  \label{table:codec_recon_all}

  \begin{tabular}{@{} c c c c c c c c c c @{}}
\toprule
\multirow{2}{*}{\textbf{Model}}
& \multirow{2}{*}{\textbf{bps}}
& \multirow{2}{*}{\makecell[c]{\textbf{Frame}\\\textbf{rate}}}
& \multirow{2}{*}{$\boldsymbol{N}_{\mathrm{VQ}}$}
& \multicolumn{4}{c}{\textbf{Speech}}
& \multicolumn{2}{c}{\textbf{Audio / Music}} \\
\cmidrule(lr){5-8}\cmidrule(lr){9-10}
& & & 
& \textbf{SIM}\,\up
& \textbf{STOI}\,\up
& \textbf{PESQ-NB}\,\up
& \textbf{PESQ-WB}\,\up
& \textbf{Mel-Loss}\,\down
& \textbf{STFT-Dist.}\,\down \\
\midrule

\textbf{StableCodec} & 700  & 25 & 2  & 0.62 / 0.45 & 0.91 / 0.86 & 2.91 / 2.50 & 2.24 / 1.93 & -- / -- & -- / -- \\
\textbf{XCodec2.0} & 800  & 50 & 1  & 0.82 / 0.74 & 0.92 / 0.86 & 3.04 / 2.46 & 2.43 / 1.96 & -- / -- & -- / -- \\
\textbf{MiMo-Audio-Tokenizer} & 850  & 25 & 4  & 0.80 / 0.74 & 0.91 / 0.87 & 2.94 / 2.62 & 2.39 / 2.14 & \bnum{0.82} / 0.81 & 2.33 / 2.23 \\
\textbf{Higgs-Audio-Tokenizer} & 1000  & 25 & 4  & 0.77 / 0.68 & 0.83 / 0.82 & 3.03 / 2.61 & 2.48 / 2.14 & 0.83 / \bnum{0.80} & 2.20 / 2.05 \\
\textbf{SpeechTokenizer} & 1000  & 50 & 2  & 0.36 / 0.25 & 0.77 / 0.68 & 1.59 / 1.38 & 1.25 / 1.17 & -- / -- & -- / -- \\
\textbf{XY-Tokenizer} & 1000  & 12.5 & 8  & 0.85 / 0.79 & 0.92 / 0.87 & 3.10 / 2.63 & 2.50 / 2.12 & -- / -- & -- / -- \\
\textbf{BigCodec} & 1040  & 80 & 1  & 0.84 / 0.69 & 0.93 / 0.88 & 3.27 / 2.55 & 2.68 / 2.06 & -- / -- & -- / -- \\
\textbf{Mimi} & 1100  & 12.5 & 8  & 0.74 / 0.59 & 0.91 / 0.85 & 2.80 / 2.24 & 2.25 / 1.78 & 1.24 / 1.19 & 2.62 / 2.49 \\
\rowcolor{oursgray} \textbf{MOSS-Audio-Tokenizer} & 750  & 12.5 & 6  & 0.82 / 0.75 & 0.93 / 0.89 & 3.14 / 2.73 & 2.60 / 2.22 & 0.86 / 0.85 & 2.21 / 2.10 \\
\rowcolor{oursgray} \textbf{MOSS-Audio-Tokenizer} & 1000  & 12.5 & 8  & \bnum{0.88} / \bnum{0.81} & \bnum{0.94} / \bnum{0.91} & \bnum{3.38} / \bnum{2.96} & \bnum{2.87} / \bnum{2.43} & \bnum{0.82} / \bnum{0.80} & \bnum{2.16} / \bnum{2.04} \\
\midrule
\textbf{DAC} & 1500  & 75 & 2  & 0.48 / 0.41 & 0.83 / 0.79 & 1.87 / 1.67 & 1.48 / 1.37 & -- / -- & -- / -- \\
\textbf{Encodec} & 1500  & 75 & 2  & 0.60 / 0.45 & 0.85 / 0.81 & 1.94 / 1.80 & 1.56 / 1.48 & 1.12 / 1.04 & 2.60 / 2.42 \\
\textbf{Higgs-Audio-Tokenizer} & 2000  & 25 & 8  & 0.90 / 0.83 & 0.85 / 0.85 & 3.59 / 3.22 & 3.11 / 2.73 & 0.74 / 0.70 & 2.07 / 1.92 \\
\textbf{SpeechTokenizer} & 2000  & 50 & 4  & 0.66 / 0.50 & 0.88 / 0.80 & 2.38 / 1.79 & 1.92 / 1.49 & -- / -- & -- / -- \\
\textbf{Qwen3-TTS-Tokenizer} & 2200  & 12.5 & 16  & \bnum{0.95} / 0.88 & \bnum{0.96} / 0.93 & 3.66 / 3.10 & 3.19 / 2.62 & -- / -- & -- / -- \\
\textbf{MiMo-Audio-Tokenizer} & 2250  & 25 & 12  & 0.89 / 0.83 & 0.95 / 0.92 & 3.57 / 3.25 & 3.05 / 2.71 & \bnum{0.70} / \bnum{0.68} & 2.21 / 2.10 \\
\textbf{Mimi} & 2475  & 12.5 & 18  & 0.89 / 0.76 & 0.94 / 0.91 & 3.49 / 2.90 & 2.97 / 2.35 & 1.10 / 1.06 & 2.45 / 2.32 \\
\rowcolor{oursgray} \textbf{MOSS-Audio-Tokenizer} & 1500  & 12.5 & 12  & 0.92 / 0.86 & 0.95 / 0.93 & 3.64 / 3.27 & 3.20 / 2.74 & 0.77 / 0.74 & 2.08 / 1.96 \\
\rowcolor{oursgray} \textbf{MOSS-Audio-Tokenizer} & 2000  & 12.5 & 16  & \bnum{0.95} / \bnum{0.89} & \bnum{0.96} / \bnum{0.94} & \bnum{3.78} / \bnum{3.46} & \bnum{3.41} / \bnum{2.96} & 0.73 / 0.70 & \bnum{2.03} / \bnum{1.90} \\
\midrule
\textbf{DAC} & 3000  & 75 & 4  & 0.74 / 0.67 & 0.90 / 0.88 & 2.76 / 2.47 & 2.31 / 2.07 & 0.86 / 0.83 & 2.23 / 2.10 \\
\textbf{MiMo-Audio-Tokenizer} & 3650  & 25 & 20  & 0.91 / 0.85 & 0.95 / 0.93 & 3.73 / 3.44 & 3.25 / 2.89 & 0.66 / 0.65 & 2.17 / 2.06 \\
\textbf{SpeechTokenizer} & 4000  & 50 & 8  & 0.85 / 0.69 & 0.92 / 0.85 & 3.05 / 2.20 & 2.60 / 1.87 & -- / -- & -- / -- \\
\textbf{Mimi} & 4400  & 12.5 & 32  & 0.94 / 0.83 & 0.96 / 0.94 & 3.80 / 3.31 & 3.43 / 2.78 & 1.02 / 0.98 & 2.34 / 2.21 \\
\textbf{Encodec} & 4500  & 75 & 6  & 0.86 / 0.75 & 0.92 / 0.91 & 2.91 / 2.63 & 2.46 / 2.15 & 0.91 / 0.84 & 2.33 / 2.17 \\
\textbf{DAC} & 6000  & 75 & 8  & 0.89 / 0.84 & 0.95 / 0.94 & 3.75 / 3.57 & 3.41 / 3.20 & \bnum{0.65} / \bnum{0.63} & 1.97 / 1.87 \\
\rowcolor{oursgray} \textbf{MOSS-Audio-Tokenizer} & 3000  & 12.5 & 24  & 0.96 / 0.92 & \bnum{0.97} / \bnum{0.96} & 3.90 / 3.64 & 3.61 / 3.20 & 0.69 / 0.66 & 1.98 / 1.84 \\
\rowcolor{oursgray} \textbf{MOSS-Audio-Tokenizer} & 4000  & 12.5 & 32  & \bnum{0.97} / \bnum{0.93} & \bnum{0.97} / \bnum{0.96} & \bnum{3.95} / \bnum{3.71} & \bnum{3.69} / \bnum{3.30} & 0.68 / 0.64 & \bnum{1.96} / \bnum{1.82} \\

    \bottomrule
  \end{tabular}
\end{table*}

\begin{figure*}[t!]
  \centering
  \includegraphics[width=0.85\linewidth]{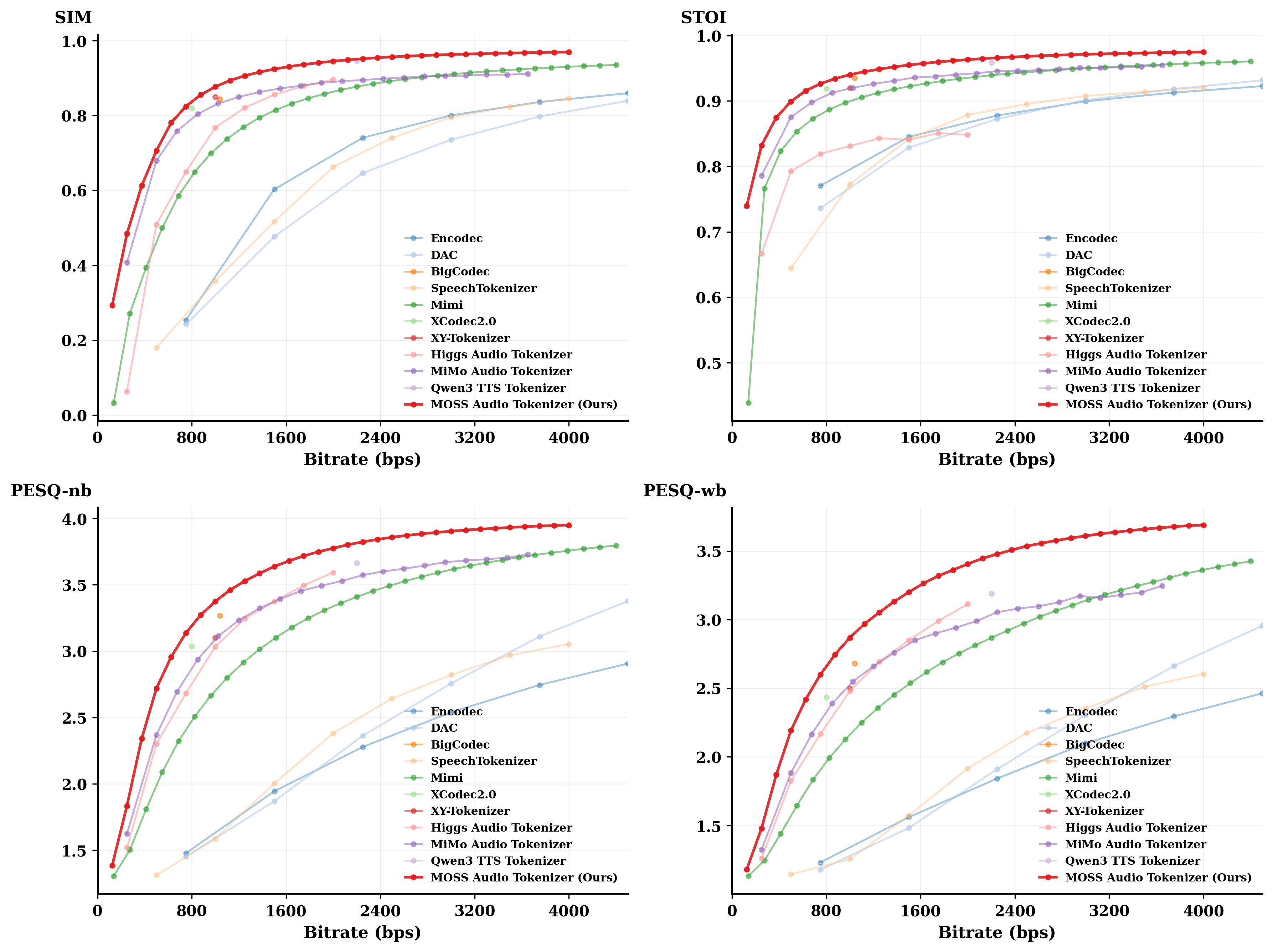}
  \caption{Comparison of objective reconstruction metrics between MOSS-Audio-Tokenizer and other state-of-the-art open-source audio tokenizers on the LibriSpeech \textit{test-clean} dataset. Results are evaluated within the 0--4\,kbps bitrate range. The horizontal axis represents the bitrate, and the vertical axis denotes the corresponding objective reconstruction scores.}
  \label{fig:metrics_librispeech}
\end{figure*}

We conduct a comprehensive evaluation of MOSS-Audio-Tokenizer, comparing it with current state-of-the-art open-source audio tokenizers across various bitrate regimes. The baseline audio tokenizers include StableCodec~\citep{parker2024scaling}, XCodec2.0~\citep{ye2025llasa}, MiMo-Audio-Tokenizer~\citep{zhang2025mimo}, Higgs-Audio-Tokenizer~\citep{higgsaudio}, SpeechTokenizer~\citep{zhang2023speechtokenizer}, XY-Tokenizer~\citep{gong2025xy}, BigCodec~\citep{xin2024bigcodec}, Mimi~\citep{defossez2024moshi}, DAC~\citep{kumar2023high}, Encodec~\citep{defossez2022high}, and Qwen3-TTS-Tokenizer~\citep{hu2026qwen3}. Our evaluation encompasses speech, general audio, and music to assess the model's versatility and reconstruction fidelity.

For speech reconstruction, we conduct evaluations on LibriSpeech test-clean (English)~\citep{panayotov2015librispeech} and AISHELL-2 (Chinese)~\citep{du2018aishell}.
We report speaker similarity (SIM), computed as the cosine similarity between speaker embeddings extracted from the original and reconstructed audio using a pretrained speaker verification model\footnote{\href{\detokenize{https://github.com/microsoft/UniSpeech/tree/main/downstreams/speaker_verification}}{UniSpeech speaker verification repository}}.
In addition, we report short-time objective intelligibility (STOI)~\citep{taal2010short} and perceptual evaluation of speech quality (PESQ)~\citep{rix2001perceptual}. 

For sound and music reconstruction, following prior work~\citep{kumar2023high}, we evaluate on the AudioSet evaluation subset~\citep{gemmeke2017audio} and MUSDB~\citep{rafii2017musdb18}.
We report mel-spectrogram distance and short-time Fourier transform (STFT) distance as objective metrics.

Table~\ref{table:codec_recon_all} summarizes the objective reconstruction results across speech, general audio, and music benchmarks. We categorize the performance into low (750--1500\,bps), medium (1500--2500\,bps), and high (2500--6000\,bps) bitrate regimes. Additionally, Figure~\ref{fig:metrics_librispeech} illustrates the performance trajectory of MOSS-Audio-Tokenizer against other open-source alternatives within the 0--4 kbps range.
Across all evaluated bitrates, MOSS-Audio-Tokenizer consistently outperforms the compared open-source baselines in speech reconstruction. On general audio and music benchmarks, the model maintains competitive performance. Notably, reconstruction quality scales gracefully with the increase in bitrate, demonstrating that the model effectively leverages additional capacity and bitrate through its joint end-to-end optimization framework.

These results indicate strong modeling capacity for MOSS-Audio-Tokenizer across both low-bitrate and high-bitrate regimes. By allowing a flexible selection of RVQ layers, the model can be adapted to diverse application requirements, spanning low-bitrate scenarios to high-fidelity audio generation. Overall, MOSS-Audio-Tokenizer provides a stable, high-fidelity, and standardized tokenizer for native audio generation models.

\subsection{Voice Cloning}
Table~\ref{tab:seed_tts_eval_voice_cloning} compares MOSS-TTS with representative open and closed systems on Seed-TTS-eval. We report both architectures in both inference modes. Results for non-MOSS baselines are collected from the corresponding technical reports and reported as given in those sources.

For prompt-conditioned generation, we distinguish two inference paradigms throughout this section. In \textbf{Clone}, the user input explicitly provides a reference audio clip. In \textbf{Continuation}, we instead prepend the reference audio to the assistant-side speech prefix, prepend its ASR transcript to the requested text, and let the model continue generating the speech for the original text. We report both modes because they probe different uses of the same pretrained model: \textbf{Clone} measures explicit reference-audio conditioning, whereas \textbf{Continuation} tests whether native speech continuation already provides usable timbre transfer without relying on a dedicated clone-style prompt format.

\begin{table*}[t!]
\centering
\small
\setlength{\tabcolsep}{3.5pt}
\caption{\textbf{Zero-shot voice cloning on Seed-TTS-eval.} We report English WER (\(\downarrow\)), English speaker similarity (SIM, \(\uparrow\)), Chinese CER (\(\downarrow\)), and Chinese SIM (\(\uparrow\)). The evaluation results are from technical reports of other models, such as VoxCPM~\citep{zhou2025voxcpm} and SparkTTS~\citep{wang2025spark}.}
\resizebox{\linewidth}{!}{
\begin{tabular}{llcccccc}
\toprule
\textbf{Model} & \textbf{Mode} & \textbf{Params} & \textbf{Open} & \textbf{EN WER} $\downarrow$ & \textbf{EN SIM} $\uparrow$ & \textbf{ZH CER} $\downarrow$ & \textbf{ZH SIM} $\uparrow$ \\
\midrule
DiTAR & -- & 0.6B & \xmark & 1.69 & 73.50 & 1.02 & 75.30 \\
FishAudio-S1 & -- & 4B & \xmark & 1.72 & 62.57 & 1.22 & 72.10 \\
CosyVoice3 & -- & 1.5B & \xmark & 2.22 & 72.00 & 1.12 & 78.10 \\
Seed-TTS & -- & -- & \xmark & 2.25 & 76.20 & 1.12 & 79.60 \\
MiniMax-Speech & -- & -- & \xmark & 1.65 & 69.20 & 0.83 & 78.30 \\
\midrule
CosyVoice & -- & 0.3B & \cmark & 4.29 & 60.90 & 3.63 & 72.30 \\
CosyVoice2 & -- & 0.5B & \cmark & 3.09 & 65.90 & 1.38 & 75.70 \\
CosyVoice3 & -- & 0.5B & \cmark & 2.02 & 71.80 & 1.16 & 78.00 \\
F5-TTS & -- & 0.3B & \cmark & 2.00 & 67.00 & 1.53 & 76.00 \\
SparkTTS & -- & 0.5B & \cmark & 3.14 & 57.30 & 1.54 & 66.00 \\
FireRedTTS & -- & 0.5B & \cmark & 3.82 & 46.00 & 1.51 & 63.50 \\
FireRedTTS-2 & -- & 1.5B & \cmark & 1.95 & 66.50 & 1.14 & 73.60 \\
Qwen2.5-Omni & -- & 7B & \cmark & 2.72 & 63.20 & 1.70 & 75.20 \\
FishAudio-S1-mini & -- & 0.5B & \cmark & 1.94 & 55.00 & 1.18 & 68.50 \\
IndexTTS2 & -- & 1.5B & \cmark & 2.23 & 70.60 & 1.03 & 76.50 \\
VibeVoice & -- & 1.5B & \cmark & 3.04 & 68.90 & 1.16 & 74.40 \\
HiggsAudio-v2 & -- & 3B & \cmark & 2.44 & 67.70 & 1.50 & 74.00 \\
GLM-TTS & -- & 1.5B & \cmark & 2.23 & 67.2 & 1.03 & 76.1 \\
GLM-TTS-RL & -- & 1.5B & \cmark & 1.91 & 68.1 & \textbf{0.89} & 76.4 \\
VoxCPM & -- & 0.5B & \cmark & 1.85 & 72.90 & 0.93 & 77.20 \\
Qwen3-TTS & -- & 0.6B & \cmark & 1.68 & 70.39 & 1.23 & 76.40 \\
Qwen3-TTS & -- & 1.7B & \cmark & \textbf{1.50} & 71.45 & 1.33 & 76.72 \\
\midrule
\multirow{2}{*}{\textbf{MOSS-TTS}} & Clone & 8B & \cmark & 1.92 & 69.31 & 1.46 & 76.21 \\
 & Continuation & 8B & \cmark & 1.84 & 70.86 & 1.37 & 76.98 \\
\midrule
\multirow{2}{*}{\textbf{MOSS-TTS-Local-Transformer}} & Clone & 1.7B & \cmark & 1.87 & 71.74 & 1.33 & 77.24 \\
 & Continuation & 1.7B & \cmark & 1.93 & \textbf{73.28} & 1.44 & \textbf{79.62} \\
\bottomrule
\end{tabular}}
\label{tab:seed_tts_eval_voice_cloning}
\end{table*}

On Seed-TTS-eval, speaker similarity is the more informative metric. Once WER/CER is already below about 2, residual differences become hard to interpret: in our manual review, most remaining mismatches in that regime are ASR errors rather than audible pronunciation failures. Under this lens, MOSS-TTS is particularly strong on SIM. For both architectures, \textbf{Continuation} consistently improves speaker similarity over \textbf{Clone}, indicating that native speech continuation is an effective way to anchor speaker identity. \textbf{MOSS-TTS-Local-Transformer} is consistently stronger than \textbf{MOSS-TTS} on speaker preservation despite using only 1.7B parameters, and \textbf{MOSS-TTS-Local-Transformer} in \textbf{Continuation} achieves the highest Chinese and English similarity scores among the open-source models in the table. This matches the architectural tradeoff discussed in Section~\ref{sec:delay_pattern} and Section~\ref{sec:local_transformer_pattern}: \textbf{MOSS-TTS-Local-Transformer} is the more modeling-efficient architecture for speaker preservation, while \textbf{MOSS-TTS} remains the simpler long-context backbone used in the control-oriented evaluations below.

\subsection{Multilingual Voice Cloning}
We evaluate the released pretrained checkpoints directly on the CV3-Eval multilingual voice cloning subset, without any task-specific fine-tuning or post-training for this benchmark. As shown in Table~\ref{tab:cv3_multilingual_voice_cloning}, this subset probes voice cloning across a larger language set than Seed-TTS-eval. We report Clone and Continuation separately for both released MOSS-TTS architectures. External baseline entries in Table~\ref{tab:cv3_multilingual_voice_cloning} are filled only where corresponding values are provided in the cited reports.

\begin{table}[t!]
\centering
\small
\setlength{\tabcolsep}{3.6pt}
\caption{\textbf{CER(\%) and WER(\%) on the CV3-Eval Multilingual Voice Cloning subset.} ``--'' means the language is unsupported.}
\begin{tabular}{llccccccccc}
\toprule
\textbf{Model} & \textbf{Mode} & \textbf{zh} & \textbf{en} & \textbf{ja} & \textbf{ko} & \textbf{de} & \textbf{es} & \textbf{fr} & \textbf{it} & \textbf{ru} \\
\midrule
F5-TTS & -- & 5.47 & 8.90 & -- & -- & -- & -- & -- & -- & -- \\
Spark-TTS & -- & 5.15 & 11.00 & -- & -- & -- & -- & -- & -- & -- \\
GPT-SoVits & -- & 7.34 & 12.50 & -- & -- & -- & -- & -- & -- & -- \\
CosyVoice2 & -- & 4.08 & 6.32 & 9.13 & 19.7 & -- & -- & -- & -- & -- \\
CosyVoice2+DiffRO & -- & 3.00 & 4.72 & 6.36 & 5.14 & -- & -- & -- & -- & -- \\
CosyVoice3-0.5B & -- & 3.89 & 5.24 & 10.4 & 12.8 & 7.41 & 4.25 & 12.9 & 6.68 & 6.77 \\
CosyVoice3-0.5B+DiffRO & -- & 2.89 & 3.68 & 5.15 & 4.02 & 4.51 & 2.99 & 8.56 & 2.94 & 3.79 \\
CosyVoice3-1.5B & -- & 3.91 & 4.99 & 7.57 & 5.69 & 6.43 & 4.47 & 11.8 & 10.5 & 6.64 \\
CosyVoice3-1.5B+DiffRO & -- & 3.01 & 3.71 & 5.27 & 4.01 & 3.93 & 3.26 & 8.09 & 2.72 & 4.11 \\
\midrule
\multirow{2}{*}{\textbf{MOSS-TTS}} & Clone & 4.42 & 4.92 & 10.72 & 6.33 & 4.70 & 4.36 & 11.17 & 5.46 & 6.37 \\
 & Continuation & 4.26 & 5.12 & 7.78 & 7.73 & 10.83 & 3.43 & 10.59 & 4.82 & 6.64 \\
\midrule
\multirow{2}{*}{\textbf{MOSS-TTS-Local-Transformer}} & Clone & 3.95 & 4.35 & 10.10 & 5.95 & 4.28 & 3.98 & 10.32 & 5.02 & 5.90 \\
 & Continuation & 3.68 & 4.89 & 7.30 & 7.20 & 10.20 & 3.10 & 9.90 & 4.40 & 6.20 \\
\bottomrule
\end{tabular}
\label{tab:cv3_multilingual_voice_cloning}
\end{table}

As shown in Table~\ref{tab:cv3_multilingual_voice_cloning}, even without benchmark-specific multilingual cloning training, MOSS-TTS remains competitive across several non-zh/en languages. Relative to strong open baselines, it shows stable performance on de/es/it/ru, and the \textbf{Continuation} setting remains usable across the broader language set despite being a harder zero-shot transfer setting. Table~\ref{tab:cv3_multilingual_voice_cloning} also shows that the largest gaps are concentrated in a few harder language pairs such as ja/ko and in some English continuation cases, which is consistent with the overall difficulty of this subset.

\subsection{Duration Control}
From this subsection onward, we report only \textbf{MOSS-TTS}. The remaining three evaluations in this section---duration control, ultra-long speech generation, and phoneme-/pinyin-level pronunciation control---stress explicit token conditioning and long-context continuation, where the delay architecture is the more practical release target because of its simpler single-backbone parameterization and better scalability at long sequence lengths. We therefore use \textbf{MOSS-TTS-Local-Transformer} primarily to characterize the similarity-quality tradeoff in the cloning benchmarks above.

We evaluate token-level duration control on \textbf{MOSS-TTS} by prompting the model with a target token count and measuring the relative duration error. Under our tokenizer, 1 second corresponds to 12.5 audio tokens. Given a target token count \(n\), the target duration is \(T_{\text{target}} = n/12.5\) seconds; we compute the realized duration \(T_{\text{real}}\) from the generated waveform and report \(\text{Err}\% = |T_{\text{real}} - T_{\text{target}}|/T_{\text{target}} \times 100\%\). We summarize errors by language and target-duration buckets.

\begin{table}[t!]
\centering
\small
\setlength{\tabcolsep}{4pt}
\caption{\textbf{Token-level duration control.} Relative duration error (\%) across target-duration buckets. AbsErr Mean: mean absolute relative error; AbsErr P50/P90: 50th/90th percentile of absolute relative error; RMSE: root mean squared relative error.}
\begin{tabular}{llcccc}
\toprule
\textbf{Language} & \textbf{Bucket} & \textbf{AbsErr Mean (\%)} $\downarrow$ & \textbf{AbsErr P50 (\%)} $\downarrow$ & \textbf{AbsErr P90 (\%)} $\downarrow$ & \textbf{RMSE (\%)} $\downarrow$ \\
\midrule
zh & 3s--10s & 1.456 & 1.333 & 2.343 & 1.652 \\
zh & 10s--1m & 0.359 & 0.254 & 0.647 & 0.502 \\
zh & 1m--10m & 0.356 & 0.077 & 1.273 & 0.849 \\
zh & 10m--30m & 0.678 & 0.061 & 1.859 & 1.228 \\
zh & overall & 0.712 & 0.284 & 2.013 & 1.141 \\
\midrule
en & 3s--10s & 1.482 & 1.357 & 2.385 & 1.685 \\
en & 10s--1m & 0.355 & 0.251 & 0.639 & 0.515 \\
en & 1m--10m & 0.365 & 0.079 & 1.304 & 0.834 \\
en & 10m--30m & 0.660 & 0.059 & 1.809 & 1.261 \\
en & overall & 0.723 & 0.288 & 2.043 & 1.160 \\
\bottomrule
\end{tabular}
\label{tab:duration_control}
\end{table}

As shown in Table~\ref{tab:duration_control}, the model achieves consistently low relative duration errors from short to long utterances, with overall AbsErr Mean around 0.7\% and strong percentile behavior. Notably, these results are obtained under a pretraining-only setup, indicating that effective token-level duration control can emerge without introducing a dedicated duration-control fine-tuning stage.

\subsection{Ultra-Long Speech Generation}
We further build an internal ultra-long evaluation set for \textbf{MOSS-TTS} to estimate expected behavior when generation extends from short utterances to approximately one hour. The set covers Chinese and English, each with six language-specific text-length buckets and 10 prompts per bucket. For each prompt, we evaluate both \textbf{Clone} and \textbf{Continuation}, yielding 240 generated utterances in total. We transcribe each sample with MOSS-Transcribe-Diarize \citep{yu2026moss} and report CER for Chinese and WER for English, each computed per sample and averaged over the 10 prompts in each bucket. Speaker similarity (SIM) is computed as the mean cosine similarity over non-overlapping 3-second windows. This internal set is used only to characterize expected performance in ultra-long generation rather than to serve as a public benchmark.

\begin{table*}[t!]
\centering
\small
\setlength{\tabcolsep}{3pt}
\caption{\textbf{Ultra-long speech generation on an internal evaluation set.} Chinese reports CER (\%) and English reports WER (\%), each averaged over 10 prompts per bucket. SIM is reported in percentage form, where 100 corresponds to perfect cosine similarity. It is computed by averaging 3-second window scores within each utterance and then averaging over utterances in the same bucket. This internal set is used only to characterize expected behavior in ultra-long generation.}
\begin{tabular}{llcccc}
\toprule
\textbf{Language} & \textbf{Bucket} & \textbf{Clone CER/WER} $\downarrow$ & \textbf{Continuation CER/WER} $\downarrow$ & \textbf{Clone SIM (\%)} $\uparrow$ & \textbf{Continuation SIM (\%)} $\uparrow$ \\
\midrule
zh & 10-100 & 0.83 & 0.65 & 69.3 & 69.9 \\
zh & 100-500 & 1.53 & 0.85 & 65.6 & 66.0 \\
zh & 500-2500 & 4.12 & 0.94 & 64.9 & 66.2 \\
zh & 2500-5000 & 3.46 & 1.19 & 63.4 & 66.3 \\
zh & 5000-10000 & 1.89 & 1.87 & 63.1 & 64.7 \\
zh & 10000+ & 3.41 & 1.86 & 60.1 & 63.0 \\
\midrule
en & 50-500 & 4.63 & 6.63 & 64.6 & 63.5 \\
en & 500-2500 & 3.65 & 4.08 & 60.9 & 60.2 \\
en & 2500-12500 & 3.75 & 4.05 & 60.3 & 60.0 \\
en & 12500-25000 & 3.76 & 3.75 & 55.5 & 56.5 \\
en & 25000-50000 & 4.58 & 6.50 & 54.8 & 53.3 \\
en & 50000+ & 17.49 & 29.52 & 44.4 & 51.2 \\
\bottomrule
\end{tabular}
\label{tab:ultra_long_internal}
\end{table*}

\begin{figure*}[t!]
  \centering
  \begin{subfigure}[t]{0.49\linewidth}
    \centering
    \includegraphics[width=\linewidth]{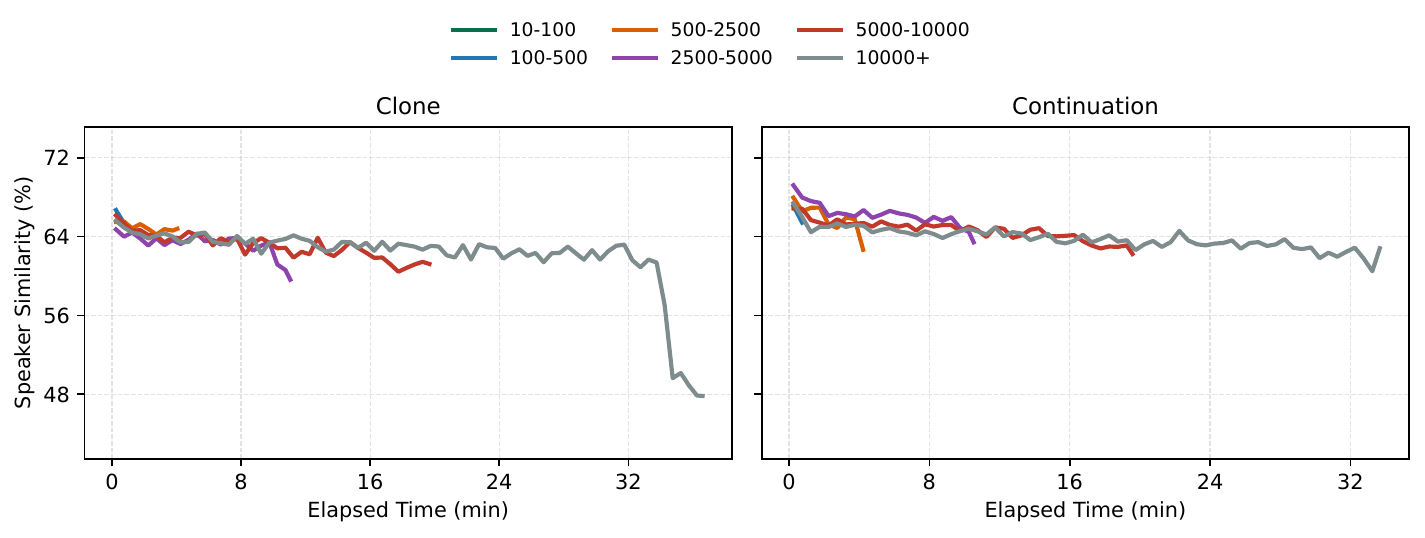}
    \caption{Chinese}
  \end{subfigure}
  \hfill
  \begin{subfigure}[t]{0.49\linewidth}
    \centering
    \includegraphics[width=\linewidth]{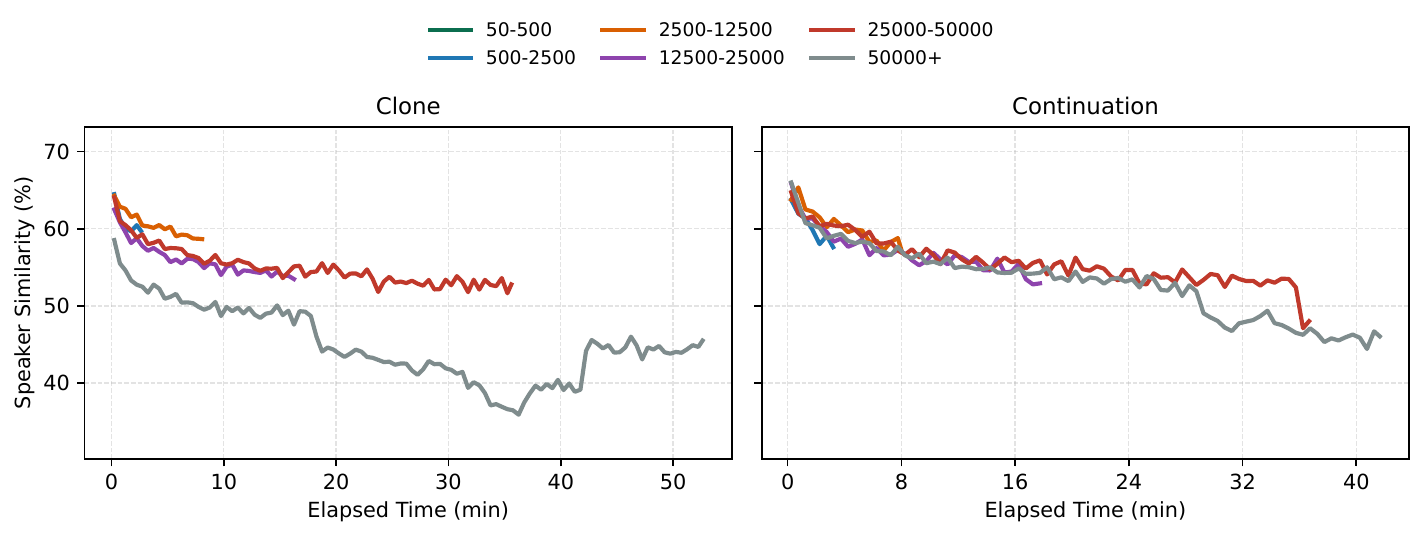}
    \caption{English}
  \end{subfigure}
  \caption{\textbf{Speaker similarity drift under ultra-long generation.} Each curve averages non-overlapping 3-second window similarities within a length bucket. For readability, the visualization reports 30-second bins and only keeps the time prefix where at least eight utterances remain in the bucket.}
\label{fig:ultra_long_sim}
\end{figure*}

Table~\ref{tab:ultra_long_internal} is best read as a coarse bucket-level summary. Content fidelity remains usable through most buckets and degrades mainly at the longest horizons, while the average SIM values already suggest that speaker preservation weakens earlier than lexical accuracy. The more informative signal, however, is the temporal drift profile in Figure~\ref{fig:ultra_long_sim}.

Figure~\ref{fig:ultra_long_sim} makes the failure mode explicit. In Chinese, most buckets begin in a narrow high-SIM band. Under \textbf{Clone}, the short and medium buckets stay fairly flat, but the 10000+ bucket shows a clear late-stage collapse near the tail. Under \textbf{Continuation}, the curves are much tighter and flatter: even the longest bucket stays close to the others for more than 30 minutes, indicating substantially better long-horizon speaker anchoring. English is harder. All buckets drift downward earlier, and the 50000+ bucket under \textbf{Clone} falls the fastest and separates from the shorter buckets after only a few minutes. \textbf{Continuation} does not remove this trend, but it clearly raises and smooths the long-bucket trajectories, especially for the 25000--50000 and 50000+ settings. The main conclusion from Figure~\ref{fig:ultra_long_sim} is therefore that ultra-long generation remains operational, but the dominant bottleneck is cumulative speaker drift over elapsed time rather than immediate lexical failure.

\begin{table}[t!]
\centering
\small
\setlength{\tabcolsep}{4pt}
\caption{\textbf{Phoneme-/pinyin-level pronunciation control on an internal evaluation set.} We report span-only CER for Chinese and span-only WER for English. Lower is better.}
\begin{tabular}{lcc}
\toprule
\textbf{Language} & \textbf{Setting} & \textbf{Replaced-Span CER/WER} $\downarrow$  \\
\midrule
zh & partial-replace & 1.00  \\
zh & full-replace & 1.65 \\
en & partial-replace & 4.32  \\
en & full-replace & 5.84  \\
\bottomrule
\end{tabular}
\label{tab:phoneme_control}
\end{table}

\subsection{Phoneme-/Pinyin-Level Pronunciation Control}
We conduct a small internal functionality evaluation for phoneme-/pinyin-level pronunciation control using \textbf{MOSS-TTS}. For each language (Chinese and English), we construct two settings: \textit{partial-replace}, where only a short target span is replaced by pinyin or IPA, and \textit{full-replace}, where the entire sentence is specified in pinyin or IPA. Each language-setting pair contains 100 samples. Since the goal of this test is to verify controllable pronunciation editing, we evaluate only the controlled span rather than the full sentence.

We transcribe each generated utterance with MOSS-Transcribe-Diarize \citep{yu2026moss}, align the transcript to the target text, and compute span-only CER (Chinese) and WER (English). As shown in Table~\ref{tab:phoneme_control}, MOSS-TTS achieves low span error in all four settings on this internal set, indicating that phoneme-/pinyin-level control is already practically usable, including both local span replacement and full-sentence phoneme control.

\section{Conclusion}
\label{sec:conclusion}

In this technical report, we presented \textbf{MOSS-TTS}, an open speech generation foundation model built on a scalable recipe: a high-quality audio tokenizer, autoregressive next-token modeling, and large-scale multilingual pretraining. Built on \textbf{MOSS-Audio-Tokenizer}, MOSS-TTS formulates speech generation as autoregressive prediction over aligned text and speech tokens. On top of this tokenizer, MOSS-TTS and MOSS-TTS-Local-Transformer instantiate two complementary operating points: the former emphasizes structural simplicity, scalability, and long-context/control-oriented deployment, while the latter emphasizes higher modeling efficiency, stronger speaker preservation, and a shorter time to first audio.

The empirical results support the central thesis of the report. MOSS-Audio-Tokenizer provides strong discrete audio tokens across bitrate regimes, and the two architectures expose a clear and practically useful tradeoff: MOSS-TTS-Local-Transformer is generally stronger on speaker similarity in zero-shot cloning, whereas MOSS-TTS is the more natural backbone for duration control and ultra-long generation. At the same time, the evaluation makes the remaining bottlenecks explicit. The hardest multilingual setting still leave room for improvement, and ultra-long generation shows that long-horizon speaker drift---rather than immediate lexical failure---is now the dominant failure mode, especially in English. We therefore view stronger long-context speaker anchoring, broader low-resource language coverage, and further improvement of fine-grained controllability as the most important next directions.

Taken together, these results suggest that speech generation can benefit from the same principles that have driven recent progress in open large language models: data quality, scale, and architectural simplicity. Rather than relying on increasingly elaborate cascades, MOSS-TTS shows that a strong tokenizer, a large-scale high-quality data pipeline, and a unified autoregressive objective already provide a practical foundation for open speech generation. With the release of MOSS-Audio-Tokenizer, MOSS-TTS, and MOSS-TTS-Local-Transformer, we hope this report can serve both as a reproducible account of the current release and as a clean baseline for future work on open speech foundation models.


\newpage
\section*{Contributors}

\noindent\textbf{Core Contributors}: \\
Yitian Gong$^{\dagger}$, Botian Jiang$^{\dagger}$, Yiwei Zhao, Yucheng Yuan, Kuangwei Chen, Yaozhou Jiang, Cheng Chang, Dong Hong, Mingshu Chen, Ruixiao Li, Yiyang Zhang, Yang Gao, Hanfu Chen, Ke Chen, Songlin Wang, Xiaogui Yang$^{*}$

\vspace{0.5em}
\noindent\textbf{Contributors}: \\
Yuqian Zhang, Kexin Huang, ZhengYuan Lin, Kang Yu, Ziqi Chen, Jin Wang, Zhaoye Fei, Qinyuan Cheng, Shimin Li

\vspace{0.5em}

\noindent\textbf{Advisors}: \\
Xipeng Qiu$^{\S}$

\vspace{1em}

\noindent\textbf{Affiliations}: \\
Shanghai Innovation Institute\\
MOSI Intelligence\\
Fudan University\\

{\let\thefootnote\relax\footnotetext{$^\dagger$Equal contribution. $^*$Project lead. $^\S$Corresponding author: \texttt{xpqiu@fudan.edu.cn}.\\We especially thank the Infrastructure and Data teams for their essential contributions to the MOSS-TTS release.}}

\clearpage
\bibliographystyle{unsrtnat}
\bibliography{main}

@article{gong2026moss,
  title={MOSS-Audio-Tokenizer: Scaling Audio Tokenizers for Future Audio Foundation Models},
  author={Gong, Yitian and Chen, Kuangwei and Fei, Zhaoye and Yang, Xiaogui and Chen, Ke and Wang, Yang and Huang, Kexin and Chen, Mingshu and Li, Ruixiao and Cheng, Qingyuan and others},
  journal={arXiv preprint arXiv:2602.10934},
  year={2026}
}

@article{dubey2024llama,
  title={The llama 3 herd of models},
  author={Dubey, Abhimanyu and Jauhri, Abhinav and Pandey, Abhinav and Kadian, Abhishek and Al-Dahle, Ahmad and Letman, Aiesha and Mathur, Akhil and Schelten, Alan and Yang, Amy and Fan, Angela and others},
  journal={arXiv preprint arXiv:2407.21783},
  year={2024}
}

@article{hu2024minicpm,
  title={Minicpm: Unveiling the potential of small language models with scalable training strategies},
  author={Hu, Shengding and Tu, Yuge and Han, Xu and He, Chaoqun and Cui, Ganqu and Long, Xiang and Zheng, Zhi and Fang, Yewei and Huang, Yuxiang and Zhao, Weilin and others},
  journal={arXiv preprint arXiv:2404.06395},
  year={2024}
}

@article{defossez2024moshi,
  title={Moshi: a speech-text foundation model for real-time dialogue},
  author={D{\'e}fossez, Alexandre and Mazar{\'e}, Laurent and Orsini, Manu and Royer, Am{\'e}lie and P{\'e}rez, Patrick and J{\'e}gou, Herv{\'e} and Grave, Edouard and Zeghidour, Neil},
  journal={arXiv preprint arXiv:2410.00037},
  year={2024}
}

@article{van2017neural,
  title={Neural discrete representation learning},
  author={Van Den Oord, Aaron and Vinyals, Oriol and others},
  journal={Advances in neural information processing systems},
  volume={30},
  year={2017}
}

@article{zhang2023speechtokenizer,
  title={Speechtokenizer: Unified speech tokenizer for speech large language models},
  author={Zhang, Xin and Zhang, Dong and Li, Shimin and Zhou, Yaqian and Qiu, Xipeng},
  journal={arXiv preprint arXiv:2308.16692},
  year={2023}
}

@article{defossez2022high,
  title={High fidelity neural audio compression},
  author={D{\'e}fossez, Alexandre and Copet, Jade and Synnaeve, Gabriel and Adi, Yossi},
  journal={arXiv preprint arXiv:2210.13438},
  year={2022}
}

@article{kumar2023high,
  title={High-fidelity audio compression with improved rvqgan},
  author={Kumar, Rithesh and Seetharaman, Prem and Luebs, Alejandro and Kumar, Ishaan and Kumar, Kundan},
  journal={Advances in Neural Information Processing Systems},
  volume={36},
  pages={27980--27993},
  year={2023}
}

@article{wang2024maskgct,
  title={Maskgct: Zero-shot text-to-speech with masked generative codec transformer},
  author={Wang, Yuancheng and Zhan, Haoyue and Liu, Liwei and Zeng, Ruihong and Guo, Haotian and Zheng, Jiachen and Zhang, Qiang and Zhang, Xueyao and Zhang, Shunsi and Wu, Zhizheng},
  journal={arXiv preprint arXiv:2409.00750},
  year={2024}
}

@article{hsu2021hubert,
  title={Hubert: Self-supervised speech representation learning by masked prediction of hidden units},
  author={Hsu, Wei-Ning and Bolte, Benjamin and Tsai, Yao-Hung Hubert and Lakhotia, Kushal and Salakhutdinov, Ruslan and Mohamed, Abdelrahman},
  journal={IEEE/ACM transactions on audio, speech, and language processing},
  volume={29},
  pages={3451--3460},
  year={2021},
  publisher={IEEE}
}

@article{li2025baichuan,
  title={Baichuan-audio: A unified framework for end-to-end speech interaction},
  author={Li, Tianpeng and Liu, Jun and Zhang, Tao and Fang, Yuanbo and Pan, Da and Wang, Mingrui and Liang, Zheng and Li, Zehuan and Lin, Mingan and Dong, Guosheng and others},
  journal={arXiv preprint arXiv:2502.17239},
  year={2025}
}

@inproceedings{ye2025codec,
  title={Codec does matter: Exploring the semantic shortcoming of codec for audio language model},
  author={Ye, Zhen and Sun, Peiwen and Lei, Jiahe and Lin, Hongzhan and Tan, Xu and Dai, Zheqi and Kong, Qiuqiang and Chen, Jianyi and Pan, Jiahao and Liu, Qifeng and others},
  booktitle={Proceedings of the AAAI Conference on Artificial Intelligence},
  volume={39},
  pages={25697--25705},
  year={2025}
}

@article{ye2025llasa,
  title={Llasa: Scaling train-time and inference-time compute for llama-based speech synthesis},
  author={Ye, Zhen and Zhu, Xinfa and Chan, Chi-Min and Wang, Xinsheng and Tan, Xu and Lei, Jiahe and Peng, Yi and Liu, Haohe and Jin, Yizhu and Dai, Zheqi and others},
  journal={arXiv preprint arXiv:2502.04128},
  year={2025}
}

@inproceedings{taal2010short,
  title={A short-time objective intelligibility measure for time-frequency weighted noisy speech},
  author={Taal, Cees H and Hendriks, Richard C and Heusdens, Richard and Jensen, Jesper},
  booktitle={2010 IEEE international conference on acoustics, speech and signal processing},
  pages={4214--4217},
  year={2010},
  organization={IEEE}
}

@inproceedings{rix2001perceptual,
  title={Perceptual evaluation of speech quality (PESQ)-a new method for speech quality assessment of telephone networks and codecs},
  author={Rix, Antony W and Beerends, John G and Hollier, Michael P and Hekstra, Andries P},
  booktitle={2001 IEEE international conference on acoustics, speech, and signal processing. Proceedings (Cat. No. 01CH37221)},
  volume={2},
  pages={749--752},
  year={2001},
  organization={IEEE}
}

@inproceedings{panayotov2015librispeech,
  title={Librispeech: an asr corpus based on public domain audio books},
  author={Panayotov, Vassil and Chen, Guoguo and Povey, Daniel and Khudanpur, Sanjeev},
  booktitle={2015 IEEE international conference on acoustics, speech and signal processing (ICASSP)},
  pages={5206--5210},
  year={2015},
  organization={IEEE}
}

@article{du2018aishell,
  title={Aishell-2: Transforming mandarin asr research into industrial scale},
  author={Du, Jiayu and Na, Xingyu and Liu, Xuechen and Bu, Hui},
  journal={arXiv preprint arXiv:1808.10583},
  year={2018}
}

@article{xin2024bigcodec,
  title={Bigcodec: Pushing the limits of low-bitrate neural speech codec},
  author={Xin, Detai and Tan, Xu and Takamichi, Shinnosuke and Saruwatari, Hiroshi},
  journal={arXiv preprint arXiv:2409.05377},
  year={2024}
}

@article{yang2025almtokenizer,
  title={ALMTokenizer: A Low-bitrate and Semantic-rich Audio Codec Tokenizer for Audio Language Modeling},
  author={Yang, Dongchao and Liu, Songxiang and Guo, Haohan and Zhao, Jiankun and Wang, Yuanyuan and Wang, Helin and Ju, Zeqian and Liu, Xubo and Chen, Xueyuan and Tan, Xu and others},
  journal={arXiv preprint arXiv:2504.10344},
  year={2025}
}

@article{kumar2019melgan,
  title={Melgan: Generative adversarial networks for conditional waveform synthesis},
  author={Kumar, Kundan and Kumar, Rithesh and De Boissiere, Thibault and Gestin, Lucas and Teoh, Wei Zhen and Sotelo, Jose and De Brebisson, Alexandre and Bengio, Yoshua and Courville, Aaron C},
  journal={Advances in neural information processing systems},
  volume={32},
  year={2019}
}

@article{zeghidour2021soundstream,
  title={Soundstream: An end-to-end neural audio codec},
  author={Zeghidour, Neil and Luebs, Alejandro and Omran, Ahmed and Skoglund, Jan and Tagliasacchi, Marco},
  journal={IEEE/ACM Transactions on Audio, Speech, and Language Processing},
  volume={30},
  pages={495--507},
  year={2021},
  publisher={IEEE}
}

@inproceedings{mao2017least,
  title={Least squares generative adversarial networks},
  author={Mao, Xudong and Li, Qing and Xie, Haoran and Lau, Raymond YK and Wang, Zhen and Paul Smolley, Stephen},
  booktitle={Proceedings of the IEEE international conference on computer vision},
  pages={2794--2802},
  year={2017}
}

@article{latif2023sparks,
  title={Sparks of large audio models: A survey and outlook},
  author={Latif, Siddique and Shoukat, Moazzam and Shamshad, Fahad and Usama, Muhammad and Ren, Yi and Cuay{\'a}huitl, Heriberto and Wang, Wenwu and Zhang, Xulong and Togneri, Roberto and Cambria, Erik and others},
  journal={arXiv preprint arXiv:2308.12792},
  year={2023}
}

@article{wu2024towards,
  title={Towards audio language modeling--an overview},
  author={Wu, Haibin and Chen, Xuanjun and Lin, Yi-Cheng and Chang, Kai-wei and Chung, Ho-Lam and Liu, Alexander H and Lee, Hung-yi},
  journal={arXiv preprint arXiv:2402.13236},
  year={2024}
}

@article{borsos2023audiolm,
  title={Audiolm: a language modeling approach to audio generation},
  author={Borsos, Zal{\'a}n and Marinier, Rapha{\"e}l and Vincent, Damien and Kharitonov, Eugene and Pietquin, Olivier and Sharifi, Matt and Roblek, Dominik and Teboul, Olivier and Grangier, David and Tagliasacchi, Marco and others},
  journal={IEEE/ACM transactions on audio, speech, and language processing},
  volume={31},
  pages={2523--2533},
  year={2023},
  publisher={IEEE}
}

@article{zhang2024speechgpt,
  title={Speechgpt-gen: Scaling chain-of-information speech generation},
  author={Zhang, Dong and Zhang, Xin and Zhan, Jun and Li, Shimin and Zhou, Yaqian and Qiu, Xipeng},
  journal={arXiv preprint arXiv:2401.13527},
  year={2024}
}

@inproceedings{radford2023robust,
  title={Robust speech recognition via large-scale weak supervision},
  author={Radford, Alec and Kim, Jong Wook and Xu, Tao and Brockman, Greg and McLeavey, Christine and Sutskever, Ilya},
  booktitle={International conference on machine learning},
  pages={28492--28518},
  year={2023},
  organization={PMLR}
}

@article{wang2023neural,
  title={Neural codec language models are zero-shot text to speech synthesizers},
  author={Wang, Chengyi and Chen, Sanyuan and Wu, Yu and Zhang, Ziqiang and Zhou, Long and Liu, Shujie and Chen, Zhuo and Liu, Yanqing and Wang, Huaming and Li, Jinyu and others},
  journal={arXiv preprint arXiv:2301.02111},
  year={2023}
}

@article{du2025cosyvoice,
  title={Cosyvoice 3: Towards in-the-wild speech generation via scaling-up and post-training},
  author={Du, Zhihao and Gao, Changfeng and Wang, Yuxuan and Yu, Fan and Zhao, Tianyu and Wang, Hao and Lv, Xiang and Wang, Hui and Ni, Chongjia and Shi, Xian and others},
  journal={arXiv preprint arXiv:2505.17589},
  year={2025}
}

@article{higgsaudio,
  title={Higgs Audio V2: Redefining Expressiveness in Audio Generation},
  author={BosonAI},
  journal={https://github.com/boson-ai/higgs-audio},
  year={2025}
}

@article{zhou2025voxcpm,
  title={Voxcpm: Tokenizer-free TTS for context-aware speech generation and true-to-life voice cloning},
  author={Zhou, Yixuan and Zeng, Guoyang and Liu, Xin and Li, Xiang and Yu, Renjie and Wang, Ziyang and Ye, Runchuan and Sun, Weiyue and Gui, Jiancheng and Li, Kehan and others},
  journal={arXiv preprint arXiv:2509.24650},
  year={2025}
}

@article{yang2025qwen3,
  title={Qwen3 technical report},
  author={Yang, An and Li, Anfeng and Yang, Baosong and Zhang, Beichen and Hui, Binyuan and Zheng, Bo and Yu, Bowen and Gao, Chang and Huang, Chengen and Lv, Chenxu and others},
  journal={arXiv preprint arXiv:2505.09388},
  year={2025}
}

@article{copet2023simple,
  title={Simple and controllable music generation},
  author={Copet, Jade and Kreuk, Felix and Gat, Itai and Remez, Tal and Kant, David and Synnaeve, Gabriel and Adi, Yossi and D{\'e}fossez, Alexandre},
  journal={Advances in Neural Information Processing Systems},
  volume={36},
  pages={47704--47720},
  year={2023}
}

@article{parker2024scaling,
  title={Scaling transformers for low-bitrate high-quality speech coding},
  author={Parker, Julian D and Smirnov, Anton and Pons, Jordi and Carr, CJ and Zukowski, Zack and Evans, Zach and Liu, Xubo},
  journal={arXiv preprint arXiv:2411.19842},
  year={2024}
}

@article{wang2025spark,
  title={Spark-tts: An efficient llm-based text-to-speech model with single-stream decoupled speech tokens},
  author={Wang, Xinsheng and Jiang, Mingqi and Ma, Ziyang and Zhang, Ziyu and Liu, Songxiang and Li, Linqin and Liang, Zheng and Zheng, Qixi and Wang, Rui and Feng, Xiaoqin and others},
  journal={arXiv preprint arXiv:2503.01710},
  year={2025}
}

@article{anastassiou2024seed,
  title={Seed-tts: A family of high-quality versatile speech generation models},
  author={Anastassiou, Philip and Chen, Jiawei and Chen, Jitong and Chen, Yuanzhe and Chen, Zhuo and Chen, Ziyi and Cong, Jian and Deng, Lelai and Ding, Chuang and Gao, Lu and others},
  journal={arXiv preprint arXiv:2406.02430},
  year={2024}
}

@article{welker2025flowdec,
  title={FlowDec: A flow-based full-band general audio codec with high perceptual quality},
  author={Welker, Simon and Le, Matthew and Chen, Ricky TQ and Hsu, Wei-Ning and Gerkmann, Timo and Richard, Alexander and Wu, Yi-Chiao},
  journal={arXiv preprint arXiv:2503.01485},
  year={2025}
}

@inproceedings{wu2023audiodec,
  title={Audiodec: An open-source streaming high-fidelity neural audio codec},
  author={Wu, Yi-Chiao and Gebru, Israel D and Markovi{\'c}, Dejan and Richard, Alexander},
  booktitle={ICASSP 2023-2023 IEEE International Conference on Acoustics, Speech and Signal Processing (ICASSP)},
  pages={1--5},
  year={2023},
  organization={IEEE}
}

@article{li2025dualcodec,
  title={Dualcodec: A low-frame-rate, semantically-enhanced neural audio codec for speech generation},
  author={Li, Jiaqi and Lin, Xiaolong and Li, Zhekai and Huang, Shixi and Wang, Yuancheng and Wang, Chaoren and Zhan, Zhenpeng and Wu, Zhizheng},
  journal={arXiv preprint arXiv:2505.13000},
  year={2025}
}

@inproceedings{du2024funcodec,
  title={Funcodec: A fundamental, reproducible and integrable open-source toolkit for neural speech codec},
  author={Du, Zhihao and Zhang, Shiliang and Hu, Kai and Zheng, Siqi},
  booktitle={ICASSP 2024-2024 IEEE International Conference on Acoustics, Speech and Signal Processing (ICASSP)},
  pages={591--595},
  year={2024},
  organization={IEEE}
}

@article{betker2023better,
  title={Better speech synthesis through scaling},
  author={Betker, James},
  journal={arXiv preprint arXiv:2305.07243},
  year={2023}
}

@article{du2024cosyvoice,
  title={Cosyvoice: A scalable multilingual zero-shot text-to-speech synthesizer based on supervised semantic tokens},
  author={Du, Zhihao and Chen, Qian and Zhang, Shiliang and Hu, Kai and Lu, Heng and Yang, Yexin and Hu, Hangrui and Zheng, Siqi and Gu, Yue and Ma, Ziyang and others},
  journal={arXiv preprint arXiv:2407.05407},
  year={2024}
}

@article{kaplan2020scaling,
  title={Scaling laws for neural language models},
  author={Kaplan, Jared and McCandlish, Sam and Henighan, Tom and Brown, Tom B and Chess, Benjamin and Child, Rewon and Gray, Scott and Radford, Alec and Wu, Jeffrey and Amodei, Dario},
  journal={arXiv preprint arXiv:2001.08361},
  year={2020}
}

@article{zhang2025mimo,
  title={MiMo-Audio: Audio Language Models are Few-Shot Learners},
  author={Zhang, Dong and Wang, Gang and Xue, Jinlong and Fang, Kai and Zhao, Liang and Ma, Rui and Ren, Shuhuai and Liu, Shuo and Guo, Tao and Zhuang, Weiji and others},
  journal={arXiv preprint arXiv:2512.23808},
  year={2025}
}

@article{hu2026qwen3,
  title={Qwen3-TTS Technical Report},
  author={Hu, Hangrui and Zhu, Xinfa and He, Ting and Guo, Dake and Zhang, Bin and Wang, Xiong and Guo, Zhifang and Jiang, Ziyue and Hao, Hongkun and Guo, Zishan and others},
  journal={arXiv preprint arXiv:2601.15621},
  year={2026}
}

@inproceedings{gemmeke2017audio,
  title={Audio set: An ontology and human-labeled dataset for audio events},
  author={Gemmeke, Jort F and Ellis, Daniel PW and Freedman, Dylan and Jansen, Aren and Lawrence, Wade and Moore, R Channing and Plakal, Manoj and Ritter, Marvin},
  booktitle={2017 IEEE international conference on acoustics, speech and signal processing (ICASSP)},
  pages={776--780},
  year={2017},
  organization={IEEE}
}

@article{rafii2017musdb18,
  title={The MUSDB18 corpus for music separation},
  author={Rafii, Zafar and Liutkus, Antoine and St{\"o}ter, Fabian-Robert and Mimilakis, Stylianos Ioannis and Bittner, Rachel},
  year={2017},
  publisher={Dec}
}

@article{gong2025xy,
  title={XY-Tokenizer: Mitigating the Semantic-Acoustic Conflict in Low-Bitrate Speech Codecs},
  author={Gong, Yitian and Jin, Luozhijie and Deng, Ruifan and Zhang, Dong and Zhang, Xin and Cheng, Qinyuan and Fei, Zhaoye and Li, Shimin and Qiu, Xipeng},
  journal={arXiv preprint arXiv:2506.23325},
  year={2025}
}

@article{dosovitskiy2020image,
  title={An image is worth 16x16 words: Transformers for image recognition at scale},
  author={Dosovitskiy, Alexey},
  journal={arXiv preprint arXiv:2010.11929},
  year={2020}
}

@article{henighan2020scaling,
  title={Scaling laws for autoregressive generative modeling},
  author={Henighan, Tom and Kaplan, Jared and Katz, Mor and Chen, Mark and Hesse, Christopher and Jackson, Jacob and Jun, Heewoo and Brown, Tom B and Dhariwal, Prafulla and Gray, Scott and others},
  journal={arXiv preprint arXiv:2010.14701},
  year={2020}
}

@article{xie2025fireredtts,
  title={Fireredtts-2: Towards long conversational speech generation for podcast and chatbot},
  author={Xie, Kun and Shen, Feiyu and Li, Junjie and Xie, Fenglong and Tang, Xu and Hu, Yao},
  journal={arXiv preprint arXiv:2509.02020},
  year={2025}
}

@article{liao2024fish,
  title={Fish-speech: Leveraging large language models for advanced multilingual text-to-speech synthesis},
  author={Liao, Shijia and Wang, Yuxuan and Li, Tianyu and Cheng, Yifan and Zhang, Ruoyi and Zhou, Rongzhi and Xing, Yijin},
  journal={arXiv preprint arXiv:2411.01156},
  year={2024}
}

@article{qwen2.5,
  title={Qwen2.5 technical report},
  author={Yang, An and Yang, Baosong and Zhang, Beichen and Hui, Binyuan and Zheng, Bo and Yu, Bowen and Li, Chengyuan and Liu, Dayiheng and Huang, Fei and Wei, Haoran and others},
  journal={arXiv preprint arXiv:2412.15115},
  year={2024}
}

@article{su2024roformer,
  title={Roformer: Enhanced transformer with rotary position embedding},
  author={Su, Jianlin and Ahmed, Murtadha and Lu, Yu and Pan, Shengfeng and Bo, Wen and Liu, Yunfeng},
  journal={Neurocomputing},
  volume={568},
  pages={127063},
  year={2024},
  publisher={Elsevier}
}

@article{oord2016wavenet,
  title={WaveNet: A Generative Model for Raw Audio},
  author={van den Oord, Aaron and Dieleman, Sander and Zen, Heiga and Simonyan, Karen and Vinyals, Oriol and Graves, Alex and Kalchbrenner, Nal and Senior, Andrew and Kavukcuoglu, Koray},
  journal={arXiv preprint arXiv:1609.03499},
  year={2016}
}

@article{wang2017tacotron,
  title={Tacotron: Towards End-to-End Speech Synthesis},
  author={Wang, Yuxuan and Skerry-Ryan, RJ and Stanton, Daisy and Wu, Yonghui and Weiss, Ron J and Jaitly, Navdeep and Yang, Zongheng and Xiao, Ying and Chen, Zhifeng and Bengio, Samy and Le, Quoc},
  journal={arXiv preprint arXiv:1703.10135},
  year={2017}
}

@article{shen2018tacotron2,
  title={Natural {TTS} Synthesis by Conditioning {WaveNet} on Mel Spectrogram Predictions},
  author={Shen, Jonathan and Pang, Ruoming and Weiss, Ron J and Schuster, Mike and Jaitly, Navdeep and Yang, Zongheng and Chen, Zhifeng and Zhang, Yu and Wang, Yuxuan and Skerry-Ryan, RJ and others},
  journal={arXiv preprint arXiv:1712.05884},
  year={2018}
}

@article{ren2019fastspeech,
  title={{FastSpeech}: Fast, Robust and Controllable Text to Speech},
  author={Ren, Yi and Ruan, Yangjun and Tan, Xu and Qin, Tao and Zhao, Sheng and Zhao, Zhou and Liu, Tie-Yan},
  journal={arXiv preprint arXiv:1905.09263},
  year={2019}
}

@article{ren2020fastspeech2,
  title={{FastSpeech} 2: Fast and High-Quality End-to-End Text to Speech},
  author={Ren, Yi and Hu, Chenxu and Tan, Xu and Qin, Tao and Zhao, Sheng and Zhao, Zhou and Liu, Tie-Yan},
  journal={arXiv preprint arXiv:2006.04558},
  year={2020}
}

@inproceedings{kim2021vits,
  title={Conditional Variational Autoencoder with Adversarial Learning for End-to-End Text-to-Speech},
  author={Kim, Jaehyeon and Kong, Jungil and Son, Juhee},
  booktitle={International Conference on Machine Learning},
  pages={5530--5540},
  year={2021},
  organization={PMLR}
}

@article{kim2020glowtts,
  title={{Glow-TTS}: A Generative Flow for Text-to-Speech via Monotonic Alignment Search},
  author={Kim, Jaehyeon and Kim, Sungwon and Kong, Jungil and Son, Juhee},
  journal={arXiv preprint arXiv:2005.11129},
  year={2020}
}

@article{popov2021gradtts,
  title={{Grad-TTS}: A Diffusion Probabilistic Model for Text-to-Speech},
  author={Popov, Vadim and Vovk, Ivan and Gogoryan, Vardan and Sadekova, Tatiana and Kudinov, Mikhail},
  journal={arXiv preprint arXiv:2105.06337},
  year={2021}
}

@article{le2023voicebox,
  title={{Voicebox}: Text-Guided Multilingual Universal Speech Generation at Scale},
  author={Le, Matthew and Vyas, Apoorv and Shi, Bowen and Karrer, Brian and Sari, Leda and Moritz, Rashel and Williamson, Mary and Manohar, Vimal and Adi, Yossi and Mahadeokar, Jay and Hsu, Wei-Ning and others},
  journal={arXiv preprint arXiv:2306.15687},
  year={2023}
}

@article{liu2023styletts2,
  title={Style{TTS} 2: Towards Human-Level Text-to-Speech through Style Diffusion and Adversarial Training with Large Speech Language Models},
  author={Li, Yinghao Aaron and Han, Cong and Raghavan, Vinay S and Mischler, Gavin and Mesgarani, Nima},
  journal={arXiv preprint arXiv:2306.07691},
  year={2023}
}

@article{casanova2022yourtts,
  title={{YourTTS}: Towards Zero-Shot Multi-Speaker {TTS} and Zero-Shot Voice Conversion for Everyone},
  author={Casanova, Edresson and Weber, Julian and Shulby, Christopher and Candido Junior, Arnaldo and G{\"o}lge, Eren and Ponti, Moacir Antonelli},
  journal={arXiv preprint arXiv:2112.02418},
  year={2022}
}

@article{tjandra2025meta,
  title={Meta audiobox aesthetics: Unified automatic quality assessment for speech, music, and sound},
  author={Tjandra, Andros and Wu, Yi-Chiao and Guo, Baishan and Hoffman, John and Ellis, Brian and Vyas, Apoorv and Shi, Bowen and Chen, Sanyuan and Le, Matt and Zacharov, Nick and others},
  journal={arXiv preprint arXiv:2502.05139},
  year={2025}
}

@article{yu2026moss,
  title={MOSS Transcribe Diarize: Accurate Transcription with Speaker Diarization},
  author={Yu, Donghua and Lin, Zhengyuan and Yang, Chen and Zhang, Yiyang and Fei, Zhaoye and Chen, Hanfu and Chen, Jingqi and Chen, Ke and Cheng, Qinyuan and Fan, Liwei and others},
  journal={arXiv preprint arXiv:2601.01554},
  year={2026}
}

@inproceedings{han2025leveraging,
  title={Leveraging self-supervised learning for speaker diarization},
  author={Han, Jiangyu and Landini, Federico and Rohdin, Johan and Silnova, Anna and Diez, Mireia and Burget, Luk{\'a}{\v{s}}},
  booktitle={Proc. ICASSP},
  year={2025}
}

@article{han2025fine,
  title={Fine-tune Before Structured Pruning: Towards Compact and Accurate Self-Supervised Models for Speaker Diarization},
  author={Han, Jiangyu and Landini, Federico and Rohdin, Johan and Silnova, Anna and Diez, Mireia and Cernocky, Jan and Burget, Lukas},
  journal={arXiv preprint arXiv:2505.24111},
  year={2025}
}

@article{han2025efficient,
  title={Efficient and Generalizable Speaker Diarization via Structured Pruning of Self-Supervised Models},
  author={Han, Jiangyu and P{\'a}lka, Petr and Delcroix, Marc and Landini, Federico and Rohdin, Johan and Cernock{\`y}, Jan and Burget, Luk{\'a}{\v{s}}},
  journal={arXiv preprint arXiv:2506.18623},
  year={2025}
}

@inproceedings{reddy2022dnsmos,
  title={{DNSMOS P.835}: A Non-Intrusive Perceptual Objective Speech Quality Metric to Evaluate Noise Suppressors},
  author={Reddy, Chandan K. A. and Gopal, Vishak and Cutler, Ross},
  booktitle={Proceedings of the IEEE International Conference on Acoustics, Speech and Signal Processing (ICASSP)},
  pages={886--890},
  year={2022},
  organization={IEEE}
}

@inproceedings{zhao2024mossformer2,
  title={{MossFormer2}: Combining Transformer and {RNN}-Free Recurrent Network for Enhanced Time-Domain Monaural Speech Separation},
  author={Zhao, Shengkui and Ma, Bin and Watanabe, Shinji},
  booktitle={Proceedings of the IEEE International Conference on Acoustics, Speech and Signal Processing (ICASSP)},
  pages={11766--11770},
  year={2024},
  organization={IEEE}
}




\end{document}